\documentclass[amsmath,amssymb,11pt]{article}
\usepackage{jheppub2}
\pdfoutput=1
\usepackage{graphicx,epsfig}
\usepackage{float}
\usepackage{subfig}
\usepackage{subfloat}
\usepackage[utf8]{inputenc}
\usepackage{hyperref}
\usepackage{caption}
\usepackage[dvipsnames]{xcolor}
\usepackage{siunitx}

\usepackage{lmodern}

\newcommand{\beq}{\begin{equation}}

\newcommand{\eeq}{\end{equation}}

 \newcommand{\be}{\begin{equation}}
 \newcommand{\ee}{\end{equation}}
 \newcommand{\bea}{\begin{eqnarray}}
 \newcommand{\eea}{\end{eqnarray}}

\usepackage{tikz}
\usetikzlibrary{positioning}
\usetikzlibrary{intersections}
\usetikzlibrary{fadings} 
\usetikzlibrary{arrows.meta} 
\usetikzlibrary{arrows}

\tikzfading[name=fade out,
inner color=transparent!0,
outer color=transparent!100]

\definecolor{cherryblossompink}{rgb}{1.0, 0.72, 0.77}
\definecolor{lightblue}{rgb}{0.68, 0.85, 0.9}

\usetikzlibrary{decorations.pathmorphing}
\usetikzlibrary{decorations.pathreplacing,decorations.markings}

\usetikzlibrary{backgrounds,automata}

\title{Diffusion in the stochastic Klein-Gordon equation}

\author{Jonathan Oppenheim,}
\author{Emanuele Panella}
\emailAdd{emanuele.panella.21@ucl.ac.uk}

\affiliation{
Department of Physics and Astronomy, University College London,\\
Gower Street, London, WC1E 6BT, United Kingdom}

\abstract{Theories of gravity in which the metric is fundamentally classical predict stochastic fluctuations in the gravitational field. In this article, we study the stochastic Klein-Gordon equation as a starting point to understand the phenomenology of linearised classical-quantum hybrid gravity. In particular, we describe how to compute the non-equilibrium two point function of the scalar field, showing explicitly the role of the initial state in regulating divergences. To do so, we use a “mod-squared-retarded'' pole-prescription and find that the covariance in the field is non-zero only outside the lightcone, scales inversely with the spatial distance of the spacetime points and grows linearly in time. The energy has a contact divergence similar to that found in the quantum case. 
We conclude by discussing possible implications of anomalous diffusion for hybrid theories of gravity, especially looking at the energy density in the predicted gravitational waves background, which can be inferred from the scalar covariances.
}

\begin{document}

\maketitle

\section{Introduction} \label{sec:intro}
A consistent theory of quantum matter and gravitation has been the central objective of modern theoretical physics since the publication of Rosenfeld's seminal paper~\cite{rosenfeld2017on-the-quantization}. A wealth of alternatives has been put forward, each with their own degree of success, in which the fundamental gravitational degrees of freedom are quantised. Yet, whilst the quantisation of low-energy perturbations of the metric around a fixed background is possible, the theory is non-renormalisable and therefore unpredictive at high energies~\cite{goroff1985ultraviolet}. Currently, we do not have consensus on a UV-complete quantum theory of gravity, nor experimental verification of gravity's quantum nature.

As such, the recent precise characterisation of theories that describe \emph{fundamentally classical} degrees of freedom interacting with quantum systems~\cite{oppenheim2021postquantum,oppenheim2022classes} -- of which some examples have been known since the mid-90's~\cite{blanchard1993on-the-interaction,diosi1995quantum,alicki2003completely,kafri2014classical,diosi2014hybrid,tilloy2016sourcing,tilloy2017principle,poulin2017information} -- opens new interesting avenues of investigations. In particular, a classical-quantum (CQ) theory of general relativity~\cite{oppenheim2021postquantum,layton2023weak,oppenheim2024diffeomorphism} can be consistently constructed, and therefore cannot be excluded in principle. Other than the interesting implications that CQ theories have for measurement theory and quantum foundations~\cite{layton2024classical,wellerdavies2024quantum,oppenheim2021postquantum,layton2025a-one-world}, CQ theories have the virtue of retaining the geometric interpretation of general relativity down to the shortest scales~\cite{oppenheim2023is-it-time,grudka2024renormalisation} and of predicting testable phenomenology that opens the doors to novel indirect tests on the quantum nature of the gravitational field~\cite{oppenheim2023gravitationally}. Indeed, hybrid theories must be irreversible~\cite{oppenheim2021postquantum,oppenheim2022classes,galley2023consistent} -- and in particular the classical sector of a CQ model evolves stochastically.

The aim of the paper is to study the classical sector of a relativistic CQ theory as a toy model which may share some phenomenology with a fundamentally classical theory of gravity. In particular, we consider the stochastic Klein-Gordon equation
\begin{equation}
\label{eq:stoch_KG_intro}
    \left(\partial_t^2-\nabla^2+m^2 \right)\phi(x) = \xi(x) \ ,\nonumber
\end{equation}
where $\xi(t)$ is a space-time white noise process, i.e. 
\begin{equation}
    \mathbb{E}[\xi(x)] = 0 \ , \qquad \mathbb{E}[\xi(x)\xi(y)] =D \  \delta^{(4)}(x-y) \ .\nonumber
\end{equation}
This is the free classical sector of a CQ field theory with a classical relativistic scalar -- for example the quartic theory presented in~\cite{oppenheim2023covariant} or the CQ Yukawa model of~\cite{grudka2024renormalisation} in which scattering has been recently discussed~\cite{carney2024classical-quantum}. While the propagating degree of freedom in General Relativity is the tensor mode, and the theory is non-linear, the study of a linear, scalar theory serves a number of purposes. First, there are scalar modes which arise in classical-quantum gravity~\cite{grudka2024renormalisation}, just as they do in related theories like quadratic gravity\cite{stelle1977renormalization}. Secondly, it serves as an interesting toy model which is useful for building up an intuition for the gravitational case. Thirdly, as we discuss in the final subsections of Section~\ref{sec:results}, the scalar result informs us also on the spectrum of the gravitational wave modes, since in transverse-traceless gauge their two polarisations evolve independently, with the amplitudes obeying a simple stochastic scalar KG equation.

In addition to analysing the stochastic wave equation as a stochastic differential equations (SDE) we will also make use of the equivalent Martin-Siggia-Rose (MSR)~\cite{martin1973statistical} path intergal representation of the stochastic Klein-Gordon field, given in Section \ref{sec:pathinte}:
\begin{equation}
    P(\phi_f,t_t) = \int\mathcal{D}\phi \mathcal{D}\tilde{\phi} e^{I_{MSR}[\phi,\tilde{\phi}]}  P(\phi_0,\dot{\phi}_0,t_0) \ , \notag
\end{equation}
where the MSR action is:
\begin{equation}
\begin{split}
    I_{MSR}[\phi,\tilde{\phi}] = \int_{\Sigma_0}^{\Sigma_f} d^4x \left[ 
    -\tilde{\phi}\left(\Box+m^2 \right)\phi +\frac{D}{2}  \tilde{\phi}^2 \right] \ . \notag
\end{split}
\end{equation}
Here, $\tilde{\phi}$ is a purely imaginary auxiliary field, known as “response variable”, which encodes the property of the noise and how it interacts with the fundamental degree of freedom $\phi$. Making use of both formulations provides a check for the main results of the article.

The objective is to build towards a complete phenomenology of stochastic gravity, in order to understand how to place bounds on the diffusion coefficient of a CQ theory of gravity from current data. Indeed, the effects of the random fluctuations can accumulate and significantly alter the deterministic gravitational state. For example, the stochasticity has been shown to be able to produce some cold dark matter (CDM) phenomenology in toy cosmological models with random evolution equations~\cite{oppenheim2024emergence}, and will lead to deviations from general relativity at low acceleration~\cite{oppenheim2024anomalous}. Other models in which stochastic fluctuations are considered is in modifications to semiclassical gravity  (e.g.~\cite{hu2008stochastic,moffat1997stochastic,ahmed2004everpresent}), with the goal of providing an effective theory of quantum gravity.  While the hypothesis of a fundamental theory constrains the form of the diffusion coefficient of the theory, which in the case where the evolution of the classical system is continuous in time,  needs to be independent of the matter degrees of freedom\cite{oppenheim2022classes}.  While stochastic fluctuations may provide an explanation for astrophysical and cosmological phenomena, the flip side of this, is that we may also be able to use astrophysical bounds to rule out a theory in which gravity remains classical. In particular, stochastic diffusion of the gravitational field could place tight constraints on the theory. We will find that anomalous diffusion in stochastic classical fields is formally identical to anomalous heating in open quantum field theory, both of which are heavily constrained by observation, and thus pose a challenge for classical-quantum theories\cite{banks1984difficulties,gross1984is-quantum,poulin2017information,oppenheim2021postquantum,oppenheim2023gravitationally,diosi2022is-there,donadi2021underground,penington2023hipshot,hotta2025classical-quantum}.

\subsection{Summary of main results}
The main result of the article is the derivation of the two-point function of the non-equilibrium Klein-Gordon field in Minkowski space, presented as Equation~\ref{eq:phiphi_m0}, showing that the second moment of the field is regular (when treated as a distribution), zero for time-like separated spacetime points, grows linearly with the total time of diffusion, and drops as $1/r$ for spacelike separated events. We find that the size of the fluctuations, and their spatial variation, is large at short distances, meaning that linearised models of CQ gravity will likely break down at small scales, where non-linearities become important -- possibly acting to smooth off the magnitude of the short-distance fluctuations. We further write down the explicit, mode-by-mode, solution for both the massless and massive field. As a result of the above derivations, we compute the energy density in the gravitational wave background for linearised CQ gravity on a flat background -- and use it to infer that the linearised regime must break down due to the size of the induced random fluctuations at macroscopic scales.

The structure of the article is as follows. We begin in Section~\ref{sec:theory} with a rapid discussion on the thermal classical Klein-Gordon field, and the damped wave equation. Not only is this a well-studied model which allows us to introduce the main concept that are key to appreciate the new technical results of this paper, it is also phenomenologically relevant in cosmological settings -- i.e. in an expanding universe a friction term arises naturally. Section~\ref{sec:results} is where the main results of our analysis are contained. We first present an argument for why the non-equilibrium correlation functions need an IR cutoff, and describe how to introduce it. We then use the weak solution to the stochastic field equation to derive the pole structure of the classical propagator -- and compute the inverse Fourier transform explicitly. Here we adopt what we call the \emph{retarded mod-squared} pole-prescription, in contrast with the Feynman mod-squared approach used in~\cite{grudka2024renormalisation} when discussing the renormalisability of CQ gravity. We compare these results with the well-known thermal correlators, and characterise the energy production in the frictionless theory.  Further, we give the explicit form for the time-evolution of the two-point function mode-by-mode in momentum space. Section~\ref{sec:discussion} concludes the paper with a discussion on what we can extract from this analysis for CQ gravity and interesting directions for future investigations. In particular, we discuss the potential need to go beyond the linear regime to describe classical perturbations of the metric about a Minkowski background, and why the non-linear terms have the potential to soften the size of the fluctuations in the gravitational field.

\subsection*{Conventions}
Unless otherwise specified, we work in natural units $4\pi G=c=\hbar=1$. 
We use 
the mostly negative 
signature for the Minkowski metric $(+---)$,
just to see if anyone reads this far.

\section{Background} 
\label{sec:theory}

In-depth knowledge of the CQ formalism is not required to follow the technical results of this article, which instead deals only with a classical stochastic system. However, since the phenomenological discussions in Section~\ref{sec:results} do rely on some of the key concepts of hybrid dynamics, we have gathered an executive summary of the main relevant ideas in Appendix~\ref{app:CQ_basics} -- where the interested reader can also find useful references to explore the topic in more detail.

\subsection{The Klein-Gordon thermal state, with friction} 
We begin by reviewing the simpler problem of the \emph{damped} stochastic wave equation:
\begin{equation}
    (\Box + m^2) \phi(x) + 3\gamma_H\dot{\phi}= \xi(x) \ ,
\label{eq:FRWish-wave}
\end{equation}
with the Gaussian random field $\xi$ (with units of inverse length squared) having the following statistics:
\begin{equation}
    \mathbb{E}[\xi(x)] = 0 \ , \qquad \mathbb{E}[\xi(x)\xi(y)] = D \  \delta^{(4)}(x-y) \ ,
\end{equation}
and $\gamma_H$ is some constant friction coefficient. The damped equation is a useful starting point, since the frictionless field -- the main object of interest for the body of this article -- is poorly understood. In the mathematics literature, this is mainly due to the fact that divergences occur in the stochastic wave equation in more than a single spatial dimensions -- meaning that a regular solution does not exist -- so deformations of the theory are usually studied instead~\cite{dalang2005hlder-sobolev,conus2008the-non-linear,balan2010the-stochastic}. Similarly, in the physics literature, the non-equilibrium system has not been studied to the best of our knowledge -- although the thermal Klein-Gordon state has been explored in detail and is now textbook material~\cite{kapusta1989finite,tong2017statistical} -- Equation~\ref{eq:FRWish-wave} is indeed the dynamics that prepares such a state.  This is why it is useful at this point to discuss the latter -- it will serve to develop an intuition to interpret the non-equilibrium results. As we will see in Section~\ref{sec:results}, the thermal field covariances are closely related to the ones of the non-equilibrium system. 

The damped dynamics, that has as a fixed-point the thermal scalar field, breaks Lorentz invariance, unless the friction term needed to flow to such a state is dynamically generated -- i.e. it is not “fundamental”. This is since the friction coefficient sets a preferred temperature and, therefore, an energy scale. The factor of 3 is there in analogy with the equations of motion of a Klein-Gordon field on an inflationary FLRW background with Hubble constant $\gamma_H$~\cite{2004}. Of course, the analogy is imperfect since in the cosmological case the d'Alembertian is the one of the FLRW geometry, whilst we consider the flat-space operator instead. In the cosmological analogy, friction can occur without a fundamental contradiction. Indeed, the loss of Lorentz invariance is natural since a preferred frame is provided by the expansion of the Universe.

For completeness, let us write these equations of motion in first order formalism by defining the conjugate momentum $\pi_\phi$ to the field
\begin{align}
\label{eq:langevin}
    \partial_t \phi &=  \ \pi_\phi\\
    \partial_t \pi_\phi &=\left(\nabla^2   - m^2 \right)\phi -3\gamma_H \pi_\phi + \xi_t \ ,
\end{align}
where:
\begin{equation}
    \mathbb{E}[\xi(t,\underline{x})\xi(t',\underline{y})] = D \delta(t-t') \delta^{(3)}(\underline{x}-\underline{y}) \ .
\end{equation}
Note that, throughout the paper, we present the stochastic differential equations (SDE) of motion in Langevin form to align with the physics literature. For the reader familiar with the theory of stochastic process, we intend any SDE in It\^o sense, with the white noise field corresponding to the formal time-derivative of a three-dimensional Brownian sheet~\cite{dalang2024stochastic}.

Because of the friction term, the system achieves a steady state. Determining that steady state is achieved by considering the evolution of the probability density $P(\phi,\pi_\phi)$ in field space. This follows from the Fokker-Planck equation (an equivalent representation of the dynamics given by Equation~\ref{eq:FRWish-wave} in terms of the evolution of the probability distribution $P$ over phase space~\cite{risken1985fokker}):
\begin{equation}
\label{eq:FP_field}
    \frac{\partial P}{\partial t} = -\int d^3 x \frac{\delta}{\delta \phi}\left(D_\phi P \right) - \int d^3 x \frac{\delta}{\delta \pi_\phi}\left(D_\pi P \right) + \frac{1}{2}\int d^3 x \int d^3 y \frac{\delta^2}{\delta \pi_\phi(x) \delta \pi_\phi(y)}\left( D_{\pi\pi} P\right) \ ,
\end{equation}
where the drift coefficients are:
\begin{align}
    D_\phi &= \pi_\phi \\
    D_\pi &= \nabla^2\phi- m^2\phi-3\gamma_H\pi_\phi  \ ,
\end{align}
whilst the diffusion coefficient is:
\begin{equation}
    D_{\pi\pi} = D \delta^{(3)}(\underline{x}-\underline{y}) \ .
\end{equation}
A natuaral ansatz for the steady state ($\partial_t P=0$) is the thermal state:
\begin{equation}
    P_T = \frac{1}{\mathcal{Z}} e^{-\beta H}\  ,
\end{equation}
with (making factors of $c$ and $\hbar$ explicit, and taking $\phi$ to be dimensionless):
\begin{equation}
\label{eq:energy}
    H = \frac{c^4}{8\pi G} \int d^3x \left(\left(\frac{4\pi G}{c^4} \right)^2\frac{\pi_\phi^2}{c^2} + (\nabla \phi)^2+  \frac{c^2m^2}{\hbar^2}\phi^2 \right) \ ,
\end{equation}
and $\beta$ to be determined ($\mathcal{Z}$ ensures normalisation on field space, whilst the numerical pre-factor is needed from dimensional analysis). Plugging this into the Fokker-Planck equation and demanding that this is a steady state imposes the Einstein relation (still keeping factors of $c$ and $\hbar$):
\begin{equation}
    \beta = \frac{24\pi  G\gamma_H}{Dc^5} \ .
\end{equation}
The steady-state distribution is a field whose modes have an average energy of $1/\beta$. Without a cutoff, the energy of the field would be divergent, as there would be infinite modes, each contributing $1/\beta$ to the total energy -- just a manifestation of the ultraviolet catastrophe. If a natural UV cutoff $\Lambda$ exists for the theory, however, the total energy depends cubically on such a scale, i.e. $H\propto \Lambda^3$ -- the volume of the physical states in reciprocal space. 

Seeing explicitly that the average energy of the thermal field corresponds to each mode contributing a $1/\beta$ is straightforward, and amounts to computing the two point functions of $\pi_\phi$ and $\phi$ in the thermal state. We now perform that computation. First, note that the probability distribution over phase space factorises between $\pi_\phi$ and $\phi$, meaning that we can compute the two separately. Let's introduce sources $J$ and $\tilde{J}$ for $\phi$ and $\pi_\phi$ respectively, defining the generating function:
\begin{equation}
\begin{split}
    Z[J,\tilde{J}] &= \int \mathcal{D}\phi \mathcal{D}\pi_\phi e^{-\beta H +\int d^3x (J\phi + \tilde{J}\pi_\phi)} \\
    &= \int \mathcal{D}\phi e^{-\frac{\beta }{2}\int d^3x \left((\nabla \phi)^2+  m^2\phi^2\right)} e^{\int d^3 x J\phi } \int \mathcal{D}\pi_\phi e^{-\frac{\beta}{2}\int d^3x \pi_\phi^2 }e^{\int d^3 x\tilde{J}\pi_\phi } \\
    &=Z_\phi[J]Z_\pi[\tilde{J}] \ .
\end{split}
\end{equation}
Performing the Gaussian integrals we obtain:
\begin{equation}
\begin{split}
    Z_\phi[J] &= \int \mathcal{D}\phi e^{-\frac{\beta}{2}\int d^3x \left((\nabla \phi)^2+ m^2 \phi^2\right)} e^{\int d^3 x J\phi } \\
    &= \int \mathcal{D}\phi e^{-\frac{\beta }{2}\int \int d^3x d^3 y \ \phi(x) \delta^{(3)}(x-y) \left(-\nabla ^2+  m^2\right)\phi(y)} e^{\int d^3 x J\phi } \\
    &= \mathcal{Z}_\phi e^{\frac{1}{2}\int \int d^3x d^3 y \ J(x) G(x-y)J(y)} 
\end{split}
\end{equation}
where $\mathcal{Z}_\phi$ is the normalisation constant of the $\phi$ probability distribution and:
\begin{equation}
  \beta\left(-\nabla ^2+ m^2\right) G(\underline{x}-\underline{y})  = \delta^{(3)}(\underline{x}-\underline{y}) \ ,
\end{equation}
i.e. as expected $G(\underline{x}-\underline{y})$ is the Green's function of the Laplacian operator with a mass term. We can easily find this in Fourier space:
\begin{equation}
    G(k) = \frac{1}{\beta} \frac{1}{k^2 +m^2} \ .
\end{equation}
The inverse Fourier transform is well known:
\begin{equation}
    G(\underline{x}-\underline{y}) = \frac{1}{4\pi \beta} \frac{1}{r} e^{-mr} =  \frac{D}{24 \pi \gamma_H} \frac{1}{r} e^{-mr}\ 
\end{equation}
and corresponds to the two-point function of the field $\phi$:
\begin{equation}
\label{eq:2point_th}
    \mathbb{E}[\phi(\underline{x})\phi(\underline{y})] = \frac{D}{24 \pi \gamma_H} \frac{1}{r} e^{-mr} \ .
\end{equation}
Differentiating in space the point-split two-point function gives the covariance for the gradient of the field at equal times:
\begin{equation}
    \mathbb{E}[\partial_i\phi(\underline{x})\partial_j\phi(\underline{y})] = \frac{D}{24 \pi \gamma_H} \left[ \left(\delta_{ij}-\frac{r_i r_j}{r^2} \right)\left(\frac{1}{r^3}+\frac{m }{r^2} \right)- \frac{r_i r_j}{r^2} \left(\frac{2}{r^3}+\frac{m}{ r^2}+\frac{m^2 }{r} \right)\right] e^{-mr} \ .
\end{equation}

We sum over all directions (recall $x^i=-x_i$ in this signature of the metric). Carefully handling the coincident limit:
\begin{equation}
\begin{split}
\label{eq:deldel_th}
    \mathbb{E}[\nabla\phi(x) \nabla\phi(y)] &= -\nabla^2 G(\underline{x}-\underline{y}) = \frac{1}{\beta} \delta^{(3)}(\underline{x}-\underline{y}) - \beta m^2G(\underline{x}-\underline{y}) \\
    & = \frac{D }{6\gamma_H} \delta^{(3)} (\underline{x}-\underline{y}) - \frac{6 \gamma_H  m^2}{D } G(\underline{x}-\underline{y}) \ .
\end{split}
\end{equation}
In the massless case, the appearance of the $\delta$-function is even more obvious: the probability distribution in terms of $\nabla\phi$ has a $\delta$-function kernel, whose inverse is the delta-function itself.

Repeating the same calculations with the momentum instead, we find that in the thermal state we have:
\begin{equation}
    \mathbb{E}[\pi(x) \pi(y)] = \frac{1}{\beta} \delta^{(3)}(\underline{x}-\underline{y}) = \frac{D }{6\gamma_H} \delta^{(3)}(\underline{x}-\underline{y}) \ .
\end{equation}
This implies that the energy of the state has a contact divergence, similarly to that found in the quantum field theory case. Indeed:
\begin{equation}
\begin{split}
\label{eq:energy_th}
\mathbb{E}[H] &= \frac{1}{2}\mathbb{E}\left[ \int d^3x \ d^3y \left(\pi_\phi(x) \pi_\phi(y) + \nabla \phi(x).\nabla\phi(y)+  m^2\phi(x)\phi(y) \right) \delta^{(3)}(\underline{x}-\underline{y})\right] \\
&=\frac{V}{\beta}\delta(0) = \frac{DV}{6\gamma_H}\delta(0) \ .
\end{split}
\end{equation}
The $\delta$-like divergence is due to the infinite number of modes contributing equally to the energy: if a cutoff scale $\Lambda$ exists, the divergence is regularised with a cubic scaling $\Lambda^3$. The total energy of the field does scale with the spatial volume $V$ -- here we regulate it with some IR cutoff which might be taken to be naturally the Hubble scale -- but of course the energy density is insensitive to the IR and only feels the contribution from the UV modes. From these two-point functions one can extend the result to covariances at unequal times by studying the eigenvalue problem of the Fokker-Planck equation~\cite{starobinsky1994equilibrium}, but we stop here. We now progress to the main results of the paper -- the covariance function of the non-equilibrium system. We will continuously refer to the thermal results for comparison.

\section{Frictionless stochastic scalar}
\label{sec:results}

We know remove friction altogether, and discuss the free stochastic wave equation. The equations of motion for the stochastic Klein-Gordon field:
\begin{equation}
\label{eq:stoch_KG}
    \left(\partial_t^2-\nabla^2+m^2 \right)\phi(x) = \xi(x) \ ,
\end{equation}
with the Gaussian random field $\xi$:
\begin{equation}
    \mathbb{E}[\xi(x)] = 0 \ , \qquad \mathbb{E}[\xi(x)\xi(y)] = D \  \delta^{(4)}(x-y) \ .
\end{equation}
Note that the diffusion coefficient is dimensionless, and the delta-function correlation is required for a local, Lorentz invariant noise.

Contrary to the damped case, the probability distribution over field space will not converge to a steady state, with the variance growing unbounded instead. This means that we cannot ignore the initial state, nor the total time of evolution. Assuming that we can initialise the system at timelike infinity, providing a Lorentz-invariant initial condition, leads to divergent results precisely for this reason. It is therefore necessary to specify an initial state on a spacelike hypersurface $\Sigma_0$, and foliate spacetime along the time-like vector specified by the initial condition. 

The necessity to specify an initial condition at finite time means that the solution will not look Lorentz-invariant, i.e. the correlators will also fail to be invariant under boosts. However, this this has nothing to do with the property of the evolution itself -- the equations of motion are perfectly Lorentz invariant. Without loss of generality -- due to linearity of the Klein-Gordon equation -- we are free to consider the initial state and its time derivative to be the identically vanishing, i.e. $\phi(t_0,\underline{x})=\dot{\phi}(t_0,\underline{x})=0$. Indeed, we can always add any solution to the (deterministic) homogeneous problem that satisfies any other initial condition. Effectively, this means we are focusing only on the deviation from the deterministic dynamics due to the stochastic fluctuations: any non-zero initial condition can be simply propagated by the deterministic equation, contributing only to a non-vanishing mean.

The weak solution to Equation~\ref{eq:stoch_KG} can be rewritten as~\cite{dalang2005hlder-sobolev}:
\begin{equation}
    \phi(x) = \int_{\Sigma_0}^{\Sigma_f}  d^4y \  G_R(x,y) \xi(y) \ ,
\end{equation}
where $G_R$ is the retarded Green's function associated to the Klein-Gordon equation. Recall that for a massive field this is given by~\cite{scharf2014finite}:
\begin{equation}
\label{eq:mass_r}
    G_R(x-y) =\Theta(x^0-y^0)\left( -\frac{1}{2\pi}\delta(\tau_{xy}^2) + \Theta(\tau_{xy}^2) \frac{m J_1(m\tau_{xy})}{4\pi \tau_{xy}}\right) \ ,
\end{equation}
where $\tau_{xy}$ is the proper time elapsed on a geodesic between $x$ and $y$, whilst $J_1$ is a Bessel function of the first kind. The propagator of the massless field trivially follows:
\begin{equation}
\label{eq:massless_r}
    G_R^0(x-y) = -\frac{\Theta(x^0-y^0)}{2\pi}\delta(\tau_{xy}^2)  \ ,
\end{equation}
and is entirely localised on the past lightcone of $x$.

As we intend the stochastic equation in the It\^o sense (i.e. the interpretation of stochastic intregration in which the noise is non-anticipative), the expectation values over realisations of the noise acts only on the random field $\xi$. Therefore:
\begin{equation}
    \mathbb{E}[\phi(x)] = \int_{\Sigma_0}^{\Sigma_f}  d^4y \  G_R(x,y) \mathbb{E}[\xi(y)] = 0 \ ,
\end{equation}
as expected. The two-point function of the field, however, is non-zero:
\begin{equation}
\begin{split}
\label{eq:cov_conv}
    \mathcal{C}(x,y|t_0) \equiv\mathbb{E}[\phi(x)\phi(y)|\phi_0=\dot{\phi}_0=0] &= \int_{\Sigma_0}^{\Sigma_f}  d^4z\int_{\Sigma_0}^{\Sigma_f}  d^4z'  \  G_R(x,z)G_R(y,z') \mathbb{E}[\xi(z)\xi(z')] \\
    &=D \int_{\Sigma_0}^{\Sigma_f}  d^4z  \  G_R(x,z)G_R(y,z)\ ,
\end{split}
\end{equation}
where we have used the fact that the random field is $\delta$-correlated in spacetime. $\mathcal{C}$ is the covariance of the field at the spacetime points $x$ and $y$ \emph{given the initial condition} of a vanishing field and conjugate momentum at the initial spacelike surface. This is valid for any stochastic field with linear equations of motion. However, performing the convolution in spacetime for a general theory is complicated -- it is much easier to go to the Fourier domain where the convolution becomes a simple multiplication, and then perform the inverse Fourier transform. For the massless KG field, solving Equation~\ref{eq:cov_conv} directly is possible, and shown in Appendix~\ref{app:spacetime_conv}. This serves as a check for the main results of this article -- the results indeed match. We now discuss solving Equation~\ref{eq:cov_conv} in Fourier space, since this method is easier to generalise. We derive the corresponding pole-prescription, which we call the \emph{mod-squared retarded} prescription,  and analyse the role of the divergent terms that appear in the computations. 

\subsection{Mod-squared retarded pole prescription}
Before considering the regularised evolution with an initial space-like surface, let's study the simpler situation in which we can extend the time integral to $\pm \infty$. The upper limit is always allowed: when computing expectation values of local observables $\mathcal{O}(x)$, we are free to extend the upper limit of the time integration to infinity -- the future evolution of the probability distribution has no bearing on the expectation value of observables at some intermediate time. Initialising the state at $-\infty$ is more problematic if the evolution does not have a fixed points, since, as mentioned before, the variance of the probability distribution in field space grows unbounded. In the case in which there exists a steady-state (e.g.~\cite{layton2025restoring}), however, extending the time integration to infinitely far away in the past would prepare such a state~\cite{panella2025classical-quantum}. 

For now, let's assume we are indeed allowed to push the lower limit of integration to time-like infinity both in the past and in the future: this will uncover the general pole-structure in Fourier space of classical correlators. Let's insert the Fourier representation of the retarded propagator in Equation~\ref{eq:cov_conv}:
\begin{equation}
\begin{split}
\label{eq:mod2}
    \mathcal{C}(x,y|t_0) &= D \int_{-\infty}^\infty dz_0 \int d^3\underline{z} \int \frac{d^4p}{(2\pi)^4}\int \frac{d^4k}{(2\pi)^4} \frac{e^{-ip(x-z)}e^{-ik(y-z)}}{\left[(p_0+i\epsilon)^2-E(\underline{p})^2 \right]\left[(k_0+i\epsilon)^2-E(\underline{k})^2 \right]} \ , \\
    & = D\int \frac{d^4p}{(2\pi)^4} \frac{e^{-ip(x-y)}}{\left[(p_0+i\epsilon)^2-E(\underline{p})^2 \right]\left[(p_0-i\epsilon)^2-E(\underline{p})^2 \right]} \ , 
\end{split}
\end{equation}
where $E(\underline{p})^2=m^2+|\underline{p}|^2$ is the relativistic energy. This structure is perfectly general for stochastic processes -- see Appendix~\ref{app:brownian} for the case of a point Brownian particle -- and we refer to it as the retarded mod-squared prescription, in contrast with the Feynman mod-squared prescription first presented in~\cite{grudka2024renormalisation} to discuss stochastic propagators. Whilst the two point function is manifestly real and symmetric under exchange of the field $\phi(x) \leftrightarrow \phi(y)$ as required for a classical two-point function for both prescriptions, the retarded one follows naturally from the weak solution to the stochastic equations. In fact, the path integral formulation of stochastic processes itself calls for the retarded prescription, in analogy with the Keldysh propagator in quantum open-system path integrals~\cite{baidya2017renormalization}, as we see in the next section.

Performing the inverse Fourier transform in Equation~\ref{eq:mod2} would lead to divergent results in the limit $\epsilon \to 0$, with the leading divergence of order $1/\epsilon$. Indeed, in the $\epsilon\to 0$ limit the 4 first-order poles become 2 second-order poles on the $p_0$ real line. Then, when computing the residue for one of them -- let's call it $P$ -- its conjugate $P^*$ will contribute with a $1/\epsilon$ factor. These are very much physical divergences -- it is not possible to map the result to distributions as it is commonly done for the spacetime representation of QFT propagators. To understand their physical origin instead, consider a Brownian particle. It is a standard result that the variance in the velocity $\dot{X}(t)$ of the particle grows linearly with time, the diffusion coefficient being the constant of proportionality -- namely $\dot{X}^2\sim D _2t$. As $t \to \infty$, the probability distribution over momenta will limit to a uniform distribution -- the variance will diverge. The same happens for the scalar undamped field when driven by a white noise process. We now will explicitly see how these divergences drop out of the correlators once an initial state is defined at finite coordinate time, and what this implies for the heating rate of the classical non-equilibrium field.

\subsection{Path integral methods}
\label{sec:pathinte}
Consider the Martin-Siggia-Rose (MSR) path-integral representation~\cite{martin1973statistical,hertz2016path} of the stochastic process given by Equation~\ref{eq:stoch_KG}. This is just the free part of the classical sector of the Yukawa CQ path-integral we outline in Appendix~\ref{app:CQ_basics}
\label{sec:classical}:
\begin{equation}
\label{eq:msr_pi}
    P(\phi_f,t_t) = \int\mathcal{D}\phi \mathcal{D}\tilde{\phi} e^{I_{MSR}[\phi,\tilde{\phi}]}  P(\phi_0,\dot{\phi}_0,t_0) \ ,
\end{equation}
where:
\begin{equation}
\begin{split}
\label{eq:msr_act}
    I_{MSR}[\phi,\tilde{\phi}] = \int_{\Sigma_0}^{\Sigma_f} d^4x \left[ 
    -\tilde{\phi}\left(\Box+m^2 \right)\phi +\frac{D}{2}  \tilde{\phi}^2 \right] 
\end{split}
\end{equation}
and $P(\phi_0,\dot{\phi}_0,t_0)$ is the probability density of the field configuration at $t=t_0$.
The two-point function of the vector $\Phi=(\phi,\tilde{\phi)}$ is, as usual, given by the inverse of the differential operator appearing in the action, subject to the boundary conditions. Here, as usual in the MSR representation of stochastic processes, we have doubled the degrees of freedom by introducing the so-called response field $\tilde{\phi}$ variable -- a purely imaginary-valued field that encodes the interaction of the physical degrees of freedom with the bath. The reader familiar with the Schwinger-Keldysh path integral representation of Lindbladian dynamics will recognise an analogy between the MSR formalism and the S-K path integral in the average-difference basis. Note, the auxiliary response fields are purely quadratic, meaning that they can be easily integrated out -- that's how one recovers the Onsager-Machlup representation of stochastic processes~\cite{onsager1953fluctuations,feynman1965quantum}. This would lead to a classical path integral over $\phi$ only, with action
\begin{equation}
    I_{OM} \propto\phi \Box^2\phi \ .
\end{equation}
Since there is a sense in which we can view the action as an equation of motion $\Box\phi=0$ squared, the OM path integral acts to suppress fluctuations away from the deterministic solution. A simple example, given in Feynman and Hibbs\cite{feynman1965quantum} is Brownian motion, with action $I_{OM}=\int dt \ddot{q}^2/2D$.

Scalar theories with such a higher-derivative action are of interest in a variety of settings -- e.g. they have been recently shown to allow for fixed UV points when coupled to gravity~\cite{boyle2025fixed} -- but it is key to highlight that, in stochastic path integrals, the action does not provide the equations of motion in the standard sense. Indeed, the Onsager-Machlup action only serves to weight the “probability” of certain configurations -- the equation of motion is still the stochastic KG equation. In fact, the Euler-Lagrange equations corresponding to such an action are 4th order in time: their solution are the most likely trajectories in field space that, given some initial conditions, generate a chosen final state~\cite{feynman1965quantum}.

For simplicity, we choose as initial condition $P(\phi_0,\dot{\phi}_0,t_0) = \delta(\phi)\delta(\dot{\phi}_0)$ -- any other initial condition can be accounted for by adding the solution to the homogeneous equation satisfying the relevant boundary conditions. In principle, one can also consider a statistical probability distribution over the initial state, but we do not consider this case here for simplicity.

As explained in the previous subsection, when computing expectation values of local observables $\mathcal{O}(x)$, we are free to extend the upper limit of the time integration to time-like infinity since the evolution is time-local and causal. The same cannot be done for the lower limit of integration. Evaluating the Green's function is then more easily computed in the Fourier domain, where the inversion of the operator is straightforward. We have:
\begin{equation}
\begin{split}
\label{eq:MSR_action}
    I_{MSR}[\phi,\tilde{\phi}] &= \int_{t_0}^\infty dx^0\int d^3x \int\frac{d^4 p}{(2\pi)^4} \frac{d^4 q}{(2\pi)^4}\left[ 
    -\tilde{\phi}(q)\left(p^2-m^2 \right)\phi(p) +\frac{D}{2}  \tilde{\phi}(q)\tilde{\phi}(q) \right] e^{-i(p+q)x} \\
    & = -\frac{1}{2}\int\frac{d^4 p}{(2\pi)^4} \frac{d^4 q}{(2\pi)^4} \Phi(-q) A(q,p) \Phi(p) \ ,
\end{split}
\end{equation}
with:
\begin{equation}
\label{eq:kinetic}
    A(q,p) =(2\pi)^3 \delta^{(3)}(\underline{p}-\underline{q}) \begin{pmatrix}
0 & p^2-m^2\\
p^2-m^2 & -D 
\end{pmatrix} \left(\pi\delta(p_0-q_0) + \frac{\mathbf{P} \ i e^{-it_0(p_0-q_0)}}{p_0-q_0}\right)
\end{equation}
where $\mathbf{P}$ indicates the principal value of the integral of Equation~\ref{eq:MSR_action}. The two-point function will then just be the inverse operator to $A$ with the appropriate boundary conditions, i.e. it will be the Green's function $G$ for such an operator -- as standard. Ignoring for a moment boundary conditions (i.e. extending the time countour to $-\infty$ for simplicity), in Fourier domain this is given by:
\begin{equation}
    G(q,p) =(2\pi)^4 \delta^{(4)}(p-q) \begin{pmatrix}
\frac{D}{(p^2-m^2)^2} & \frac{1}{p^2-m^2}\\
\frac{1}{p^2-m^2} & 0 
\end{pmatrix} \ .
\end{equation}
Of course, this must be accompanied by the appropriate recipe to go around poles in the complex plane for all of these propagators. In MSR, response variables have causal dependence with the real field, meaning that the $\phi \tilde{\phi}$ correlator is of retarded form (it is non-zero only if the field \emph{precedes} the response variable). This means that, naturally, the MSR path integral implies the mod-squared retarded prescription for the diagonal $\phi \phi$ propagator, in agreement with what we derived in the previous section.

\subsection{Correlators in position space}
Here, we are only interested in the explicit form of the $\phi\phi$ correlators -- as the $\tilde{\phi}\phi$ is already known (it is just the retarded propagator of a scalar field). The subtleties of inverting in Fourier domain Equation~\ref{eq:kinetic} for finite $t_0$ can be therefore sidestepped easily by looking at Equation~\ref{eq:cov_conv} directly, and just performing the inverse Fourier transform -- which is what we do now. Let's insert the Fourier representation of the retarded propagator in Equation~\ref{eq:cov_conv}:
\begin{equation}
    \mathcal{C}(x,y|t_0) = D \int_{t_0}^\infty dz_0 \int d^3z \int \frac{d^4p}{(2\pi)^4}\int \frac{d^4k}{(2\pi)^4} \frac{e^{-ip(x-z)}e^{-ik(y-z)}}{\left[(p_0+i\epsilon)^2-E(\underline{p})^2 \right]\left[(k_0+i\epsilon)^2-E(\underline{k})^2 \right]} \ .
\end{equation}
Now, let's focus on the $z_0$ integration first:
\begin{equation}
    \int_{t_0}^\infty dz_0 e^{iz_0(p_0+k_0)} = e^{i(p_0+k_0)t_0} \left(\pi \delta(p_0+k_0)+ i \frac{\mathbf{P}}{k_0+p_0} \right) ,
\end{equation}
It would be tempting to use at this point the Kramers-Kronig relation
\begin{equation}
    \mathbf{P}\frac{1}{x} = \lim_{\epsilon\to 0}\frac{1}{x+i\epsilon}+i\pi\delta(x)
\end{equation}
and remove a delta-function. Unfortunately, this would lead down the line to a $p_0$ pole in the integrand without a definite $i\epsilon$ prescription, so we refrain from doing it and evaluate the two components separately. The spatial $z$ integral trivially gives a delta-function on momenta:
\begin{equation}
    \int d^3z e^{-iz(p+k)} = (2\pi)^3 \delta(\underline{p}+\underline{k}).
\end{equation}
Performing the integrals over $k$ and combining we obtain (see Appendix~\ref{app:four_int} for more details):
\begin{equation}
\mathcal{C}(x,y|t_0) = \mathcal{C}_\infty + \Delta\mathcal{C},
\end{equation}
where
\begin{equation}
\label{eq:infinte}
    \mathcal{C}_\infty
    = D \int \frac{d^4p}{(2\pi)^4} \frac{e^{-ip(x-y)}}{\left[(p_0+i\epsilon)^2-E(\underline{p})^2 \right]\left[(p_0-i\epsilon)^2-E(\underline{p})^2 \right]}  \ ,
\end{equation}
is the infinite-time contribution, whilst
\begin{equation}
\label{eq:finite}
    \Delta\mathcal{C} = \frac{D}{2} e^{-\epsilon(y_0-t_0)} \int \frac{d^4p}{(2\pi)^4}\frac{1}{E(\underline{p})}\frac{e^{-i\underline{p}.\underline{y}}e^{-ipx}e^{ip_0 t_0}}{(p_0+i\epsilon)^2-E(\underline{p})^2} \left[\frac{e^{-iE(\underline{p})(y_0-t_0)}}{p_0-i\epsilon+E(\underline{p})} -  \frac{e^{iE(\underline{p})(y_0-t_0)}}{p_0-i\epsilon-E(\underline{p})}\right] \ 
\end{equation}
contains information about finite-times effects.

The integral in Equation~\ref{eq:infinte} has a leading $1/\epsilon$ divergence. However, so does Equation~\ref{eq:finite} -- in such a way that the two divergent contributions cancel and one is left with a finite term only. Again, details are in Appendix~\ref{app:four_int}. We report directly the finite result in terms of oscillatory integrals, namely:
\begin{equation}
\label{eq:sol_int}
    \mathcal{C}(x,y|t_0) = \frac{D}{8\pi^2 } \frac{1}{|\underline{x}-\underline{y}|} \left[2(y^0-t_0) I_1 + I_2 - I_3  \right] \ ,
\end{equation}
with the oscillatory integrals
\begin{align}
    I_1 &= \int_0^\infty dp \frac{p}{E(p)^2} \cos\left(E(p) (x^0-y^0) \right) \sin(p|\underline{x}-\underline{y}|) \ , \\
    I_2 &= \int_0^\infty dp \frac{p}{E(p)^3} \sin\left(E(p) (x^0-y^0) \right) \sin(p|\underline{x}-\underline{y}|) \ , \\
    I_3 &= \int_0^\infty dp \frac{p}{E(p)^3} \sin\left(E(p) (x^0+y^0-2t_0) \right) \sin(p|\underline{x}-\underline{y}|) \ ,
\end{align}
where $I_3$ is really the same as $I_2$ -- only with different time coordinates. While Equation~\ref{eq:sol_int} seems to give different weighting to $x^0$ and $y^0$, this is a residue of the fact that in the calculations we have assumed $x^0 \geq y^0$ throughout. To get the general expression, it suffices to replace $x^0 \to \mathrm{Max}(x^0,y^0)$ and  $y^0 \to \mathrm{Min}(x^0,y^0)$. It further turns out that evaluating the integrals explicitly in the massless case yields a result that is invariant under $x^0-y^0 \to y^0-x^0$.

\subsubsection{Massless field}
In the massless case, the integrals simplify greatly and can be computed exactly. This is the case of phenomenological relevance, if we want to think about the scalar waves as a proxy for the tensor gravitational waves. Even if they do acquire a small mass due to renormalisation effects, as suggested in~\cite{grudka2024renormalisation}, the corrections to the massless result would only act to suppress long-distance correlations. As we show explicitly in Appendix~\ref{app:four_int}, the 2-point function for the massless field is given by:
\begin{equation}
\label{eq:phiphi_m0}
    \mathcal{C}(x,y|t_0) =\frac{D}{16\pi } \left( \frac{y^0+x^0-2t_0}{|\underline{x}-\underline{y}|} -1\right)\Theta(-s_{xy}^2) \Theta(x^0+y^0-2t_0- |\underline{x}-\underline{y}|)\ ,
\end{equation}
which indeed matches the result obtained by convolution of the Green's function, computed in Appendix~\ref{app:spacetime_conv}. Equation~\ref{eq:phiphi_m0} is the main technical result of this article. It is finite and well-behaved modulo the contact divergence at $x\to y$, that is to be handled as usual in field theory -- as a distribution acting on regular functions. The IR divergence that we have removed with the definition of the initial state at finite coordinate time $t_0$ is recovered as we send $t_0\to-\infty$.

\begin{figure}[ht]
\centering
\begin{tikzpicture}[scale=1.0, every node/.style={font=\small}]

\draw[->, thick] (-4.5,0) -- (4.5,0) node[anchor=west] {\(x\)};
\draw[->, thick] (0,-0.5) -- (0,6) node[anchor=south] {\(t\)};

\draw[dashed, thick] (-4,0.5) -- (4,0.5);

\node[above right] at (0,0.5) {\(t_0\)};

\coordinate (A) at (-4,0);
\coordinate (B) at (-2,2);
\coordinate (C) at (0,0);
\draw[thick, red] (A) -- (B) -- (C);
\node[red] at (-2,2.3) {\(x\)};
\filldraw[red] (B) circle (2pt);

\coordinate (Y1) at (1.5,1);
\coordinate (Y1L) at (0.5,0);
\coordinate (Y1R) at (2.5,0);
\draw[thick, blue] (Y1L) -- (Y1) -- (Y1R);
\node[blue] at (1.5,1.3) {\(y\)};
\filldraw[blue] (Y1) circle (2pt);

\coordinate (Y2) at (2.5,3.5);
\coordinate (Y2L) at (-1.5,0);
\coordinate (Y2R) at (3.5,2.5);
\draw[thick, blue] (Y2L) -- (Y2) -- (Y2R);
\node[blue] at (2.5,3.8) {\(\ y'\)};
\filldraw[blue] (Y2) circle (2pt);

\coordinate (INT) at (-0.7,0.7);
\filldraw[black] (INT) circle (2pt);

\end{tikzpicture}
\caption{Spacetime diagram in \(1+1\) dimensions. The points $x$ and $y$, even if spacelike separated, are not correlated -- their lightcones intersect \emph{before} the initial condition (dashed line). On the other hand, the field at $x$ and $y'$ will have non-zero covariance, as their intersection (black dot) lies in the future of the initial spacelike hypersurface. \label{fig:spacetime_diag}
}
\end{figure}
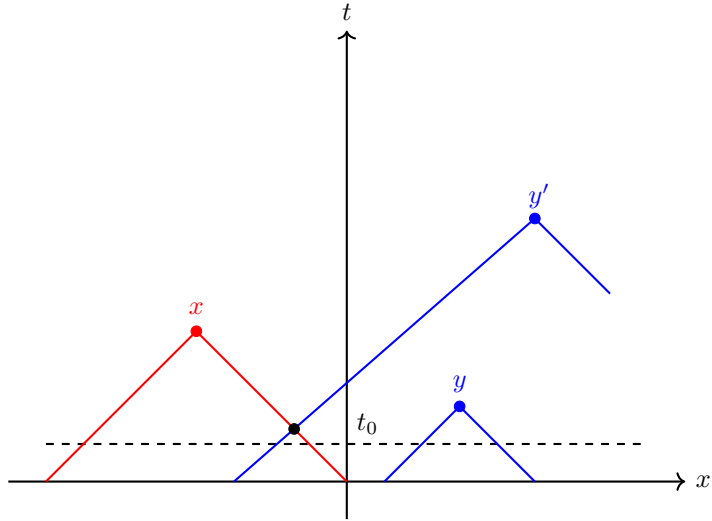

The fluctuations in the field at time-like separated points are un-correlated. This is caused by the structure of the massless retarded propagator -- which is completely localised on the lightcone. In order for the field in two different spacetime points to be correlated, they need to share signal from the same stochastic fluctuation. Since these travel strictly at the speed of light, this implies that $x$ and $y$ need to have intersecting light-cones -- hence the first $\Theta$-function. The second $\Theta$-function follows from the initial condition $\phi=0$ for $t<t_0$. Only if the $\Theta$-condition is satisfied, then the two lightcones intersect before the stochastic white noise is turned on, i.e. before $t_0$. Indeed, the covariance between $x$ and $y$ drops as $1/r$ until it reaches zero at a critical distance $r_*$. Further than $r_*$, the two space-like separated points do not have an intersecting lightcone and are completely uncorrelated. Figure~\ref{fig:spacetime_diag} visually clarifies this point. Note that the critical distance $r_*$ grows linearly with the time coordinate elapsed from the initial spacelike hypersurface.

A useful analysis, for a further sanity check, is to compare Equation~\ref{eq:phiphi_m0} for simultaneous events (with respect to the initial time hypersurface) to Equation~\ref{eq:2point_th}, i.e. the thermal 2-point function. By taking the time elapsed to be $t_0 \sim 1/3\gamma_H$, i.e. the thermalisation scale, we see that in the massless limit both the $1/r$ scaling and size of the correlations match. Of course, for the thermal state the $\Theta$-function depending on the initial time is absent -- in the equilibrium configuration enough time has elapsed such that all space-like separated points have intersecting lightcones.

\subsection{Energy production}
While in an expanding universe, we expect Hubble friction terms such as those found in \eqref{eq:FRWish-wave}, we here consider Minkowski space without additional friction terms. In this case, we expect the energy of the system to increase with time, in analogy with a simple Brownian particle. Since the Hamiltonian is not the generator of the dynamics there is no reason for it to be conserved. The theory is time-translation invariant, but Noether's theorem doesn't apply.

It is straightforward to show this diffusion in energy by referring to It\^o's lemma. In particular, given a function $f(z)$ of a stochastic variable $z$ that obeys an It\^o SDE of the form
\begin{equation}
    \dot{z} = \mu(z)+\sigma(z)\xi(t) \ ,
\end{equation}
then we have:
\begin{equation}
    \dot{f} = \mu(z)\frac{\partial f}{\partial z}+\frac{\sigma^2}{2} \frac{\partial^2f}{\partial z^2}   + \sigma(z)\frac{\partial f}{\partial z} \xi(t)\ .
\end{equation}
The term involving second derivatives is an extra contribution to the standard chain rule, arising from the fact that, pictorially, the white noise field has fractional time dimensions $\xi \sim dt^{-1/2}$. This naturally generalises to multivariate processes and, by extension, field theories.

Let's now consider the observable in Equation~\ref{eq:energy}, the energy $E$ of the field. By applying It\^o's lemma we obtain:
\begin{equation}
    \dot{E} = \int d^3x \left(\sqrt{D} \pi(x) \xi(x) + \frac{1}{2}D \delta(0) \right) \ ,
\end{equation}
where the deterministic contributions to the chain rule have cancelled each others (the Klein-Gordon equation itself is energy-preserving), and only diffusion effects survive. Using that the expectation value of the noise process vanishes, we finally obtain that the energy increases linearly with time in a deterministic fashion:
\begin{equation}
    \Delta E = \frac{Dc^5V}{8\pi G}\delta(0) \Delta t \ ,
\label{eq:energyincrease}
\end{equation}
where we have reintroduced factors of $c$ and $\Delta t$ is the time elapsed from the initial condition (recall, $D$ is dimensionless). Again, this matches Equation~\ref{eq:energy_th}, the thermal state result. Much like for the energy of the thermal state, the $\delta$-divergence is due to the infinite number of modes in the system that are being excited. The energy of the quantum scalar field has a similar contact divergence, suggesting that this is equivalent to the cosmological constant problem. In the case of effective theories with a natural breakdown scale, it can be simply dealt with by introducing a UV cutoff $\Lambda$.
We discuss possible implications for fundamental CQ field theories in the Discussion section.

A note of caution. Naively, one would expect that each mode with spatial momentum $p$ contributes with $1/p^2$ to the total energy of the field at a certain fixed $t$ time slice, the two-point function $\phi^2$ in momentum space scales like $1/p^4$. However (assuming for simplicity that the total energy of the field equipartitions between the time and spatial derivatives, as it does in the thermal state):
\begin{equation}
\begin{split}
    \mathbb{E}[H(\Delta t)] &\sim \int d^3x \ \mathbb{E}[\nabla\phi(\underline{x}).\nabla\phi(\underline{x})] \\
    &\sim \int d^3p \ p^2 \mathbb{E}[\phi(p,\Delta t)\phi(-p,\Delta t)]\\
    &\sim \int d^3p \ p^2 \frac{\Delta t}{p^2} \ .
\end{split}
\end{equation}
From the second to the third line, we have used the leading $\Delta t$ contribution to the two-point function of the field \emph{after} integrating over $p_0$, given the initial state (see Appendix~\ref{app:four_int}). Hence, each mode will contribute equally to the energy, with an amplitude that increases linearly with the time of diffusion. The total energy of the field relates therefore to the total volume of reciprocal space spanned by the modes of the theory. Indeed, assuming a cutoff energy scale $\Lambda$, we have:
\begin{equation}
    \Delta E= \frac{Dc^5V}{8\pi G} \Lambda^3 \Delta t \ .
\end{equation}

A distinct feature of this anomalous diffusion is that the classical energy density scales as $\Delta t\Lambda^3$, whilst the Lorentz invariant QFT vacuum zero-point energy scales as $\Lambda^4$. 

\subsubsection{Comparison to the quantum case}
The divergent energy production we observe in the classical stochastic field theory shares a lot of similarities with its better-known counterpart, the scalar open quantum field theory. To see this, consider the following representation of the Fokker-Planck evolution in Equation~\ref{eq:FP_field} (with no friction) for the probability density $P$ in configuration space:
\begin{align}
    \frac{\partial P}{\partial t}=\{H,P\}+\frac{D} {2}\int d^3x\{\phi,\{\phi,P\}\}.
\end{align}
It can be used to compute the diffusion in energy via the “Heisenberg representation”
\begin{align}
\frac{dH}{d t}=\frac{D} {2}\int d^3x\{\phi,\{\phi,H\}\}
\end{align}
which gives the linear in time energy increase proportional to the contact divergence
found in Equation~\ref{eq:energyincrease}. Whilst this is just a different representation of It\^o's lemma, it is cosmetically very similar to the computation in the quantum case. 

Indeed, a similar contact divergence appears in the energy production of the quantum scalar field, owing to the presence of modes with arbitrarily high energy. To see this, consider the following Lindblad equation for scalar fields
\begin{align}
    \frac{\partial\hat{\sigma}}{\partial t}=-i\left[\hat{H},\hat{\sigma}\right]-\frac{D_0} {2}\int d^3xd^3x'h(x,x')\left[\hat{\phi}(x),\left[\hat{\phi}(x'),\hat{\sigma}\right]\right]
\end{align}
for the density matrix $\hat{\sigma}$ and Lindblad operators given by the scalar field operators $\phi$~\cite{banks1984difficulties,ghirardi1990relativistic,baidya2017renormalization,oppenheim2021postquantum,diosi2022is-there}. This has been studied in the context of open quantum field theory, i.e. in the case where the generator of the evolution is not the Hamiltonian, but a Lindbladian with field-dependent decoherence operators. The similarity with the classical evolution is beyond cosmetic -- in fact when the Lindblad operators are linear in the field, the quantum dynamics can be mapped exactly to our classical diffusive evolution in a similar fashion as what done in~\cite{panella2025classical-quantum,panella2025new} for the coupled hybrid oscillator.  In particular, when computing the evolution of the Hamiltonian under Markovian, Lindbladian evolution, and “ultra-local” coupling $h(x,x')\rightarrow\delta^{(3)}(x-x')$,  one obtains a similar contact divergence in the heating rate to the one  found in~\cite{banks1984difficulties}. In the quantum field theory this is famously problematic, since it corresponds to an infinite production rate of bosons,
as can be seen by deriving the time evolution for the bosonic number operator in the Heisenberg representation of the Lindblad dynamics. Of course, although the energy production in this model may be inconsistent with observation, it does not violate any law of physics -- the generator of the dynamics for this QFT model is specifically not the Hamiltonian, and the theory is therefore non-unitary, although it is completely positive trace-preserving (CPTP). So, time translation invariance doesn't imply the conservation law -- Noether's theorem doesn't apply. 

\subsection{Exact mode solution}
Whilst the path integral techniques discussed here are very general and can be applied to a variety of settings, the evolution of the probability distribution for the stochastic Klein-Gordon field can be computed exactly, mode by mode. To see this, introduce the following Fourier mode decomposition for the field and its conjugate momentum
\begin{equation}
\label{eq:mode_phipi}
    \phi(\underline{x},t) = \int \frac{d^3 k}{(2\pi)^3} \phi_k(t) e^{i \underline{k}.\underline{x}} \ , \qquad \pi_{\phi}(\underline{x},t) = \int \frac{d^3 k}{(2\pi)^3} \pi_k(t) e^{i \underline{k}.\underline{x}} \ ,
\end{equation}  
where for ease of notation we defined $f(\underline{k},t) \equiv f_k(t)$. Similarly, let's introduce the mode expansion for the noise field
\begin{equation}
\label{eq:mode_xi}
    \xi(\underline{x},t) = \int \frac{d^3 k}{(2\pi)^3} \xi_k(t) e^{i \underline{k}.\underline{x}} \ ,
\end{equation}
it is straightforward to see that the noise moments imply
\begin{equation}
    \mathbb{E}[\xi_k(t) \xi_{k'}(t')] = D \delta(t-t') \delta^{(3)}(\underline{k}+\underline{k}') \ .
\end{equation}
Then, plugging in Equations~\ref{eq:mode_phipi} and~\ref{eq:mode_xi} into the first order (frictionless) stochastic equations given by Equation~\ref{eq:langevin}, we find that the equations neatly separate into  mode for mode, giving
\begin{equation}
\label{eq:mode_lang}
    \dot{\phi}_k = \pi_k \qquad \dot{\pi}_k = - E_k^2 \phi_k + \xi_k \ .
\end{equation}
These are a tower of simple, independent, two-dimensional (degenerate) Ornstein-Ulhenbeck (O-U) processes. O-U processes are Gaussian-preserving, and the evolution of the moments of the probability distribution can be computed exactly. For simplicity, we take the mean to be zero, since it can be easily re-introduced (it just follows a solution to the Klein-Gordon equation with appropriate initial conditions). Then, consider each independent mode to be sampled from an initial mean-zero Gaussian distribution at $t_0=0$(we have assumed throughout that the initial state is a $\delta$-function, which can be trivially recovered as its zero-variance limit)
\begin{equation}
    P_k(\Phi_k,0) = \frac{1}{2\pi\sqrt{\mathrm{det[}C_k(0)]}}\exp\left(-\frac{1}{2}\Phi_k^T C^{-1}_{k}(0)\Phi_k\right) \ .
\end{equation}
The evolution equation preserve the mean-zero \emph{and} the Gaussianity of the state, meaning that the only non-trivial degree of freedom to be solved for is the covariance matrix for the field and momentum of each mode. Note that here we have introduced the “field vector” $\Phi_k=(\phi_k,\pi_k)$. It is a standard result that for a O-U process
\begin{equation}
    d\Phi_k = \Theta_k \Phi_k \ dt + \Sigma \ dW_t \ ,
\end{equation}
where $\Theta$ and $\Sigma$ are constant (field-independent) matrices, the covariance of the mean-subtracted process evolves following the time-dependent Lyapunov equation:
\begin{align}
    dC_k &= d\mathbb{E}[\Phi_k \Phi_k^T] = d\mathbb{E}[d\Phi_k \Phi_k^T + \Phi_kd\Phi_k^T + d\Phi_kd\Phi_K^T] \\
    &=(\Theta_k C_k +C_k \Theta^T_k + \Sigma\Sigma^T)dt \ ,
\end{align}
where the $(d\Phi_k)^2$ term has to be kept in the variation by It\^o's lemma. By matching with the mode equations of motion in Equation~\ref{eq:mode_lang}, we can see that the evolution equations for the elements of the covariance matrix are given by
\begin{align}
    \dot{C}_{k,\phi\phi} &= 2C_{k,\pi,\phi} \\
    \dot{C}_{k,\pi\pi} &= -2E_k^2C_{k,\pi,\phi} + D \\
    \dot{C}_{k,\pi\phi} &= C_{k,\pi,\pi}-E_k^2 C_{k,\phi\phi} \ ,
\end{align}
which can be solved exactly
\begin{equation}
\begin{split}
\label{eq:c_kphiphi}
    C_{k,\phi\phi}(t) = & \frac{C_{k,\phi\phi}(0)}{2} \left( 1+\cos(2E_kt)\right) +\frac{C_{k,\pi\pi}(0)}{2 E_k^2}  \left( 1-\cos(2E_kt)\right) \\
    & \qquad + \left( C_{k,\pi\phi}(0)-\frac{D}{4E_k^2}\right)\sin(2E_kt) +\frac{D}{2E_k^2}t \ ,
\end{split}
\end{equation}
\begin{equation}
\begin{split}
    C_{k,\pi\pi}(t) = & \frac{C_{k,\pi\pi}(0)}{2} \left( 1+\cos(2E_kt)\right) +\frac{E_k^2 C_{k,\phi\phi}(0)}{2}  \left( 1-\cos(2E_kt)\right) \\
    & \qquad - \left( E_kC_{k,\pi\phi}(0)-\frac{D}{4E_k}\right)\sin(2E_kt) +\frac{D}{2}t \ ,
\end{split}
\end{equation}
\begin{equation}
\begin{split}
    C_{k,\pi\phi}(t) = &  \left( C_{k,\pi\phi}(0)-\frac{D}{4E_k^2}\right)\cos(2E_kt) - \frac{1}{2}\left( C_{k,\phi\phi}(0)-\frac{C_{k,\pi\pi}(0)}{E_k}\right)\sin(2E_kt) \\
    & \qquad + \frac{D}{4 E_k^2} \ .
\end{split}
\end{equation}
As per the now clear theme of these work, variances grow linearly in time mode by mode, both for the field and their conjugate momenta -- and therefore so does the energy of the Klein-Gordon solution. By combining the evolution of the probability distribution mode per mode it is straightforward to reconstruct the time evolution of every initial Gaussian probability distribution over field space.

\subsection{Implications for CQ gravity}\label{sec:pheno}
The last few years have seen a surge in proposals for low-energy signatures of quantum gravity. Current proposals include measuring gravitationally-induced entanglement~\cite{bose2017spin,marletto2017gravitationally}, quantum-induced noise in the gravitational field~\cite{pang2018quantum,parikh2020the-noise} (which may be measurable for highly squeezed states~\cite{carney2024response}) and others~\cite{lami2024testing,carney2024graviton,kryhin2025distinguishable}. However, these alternative tabletop experiments still require some significant technological developments ~\cite{carney2019tabletop}. 

A condition for hybrid theories to be consistent (and to escape various no-go theorems~\cite{bergmann1957summary,eppley1977tnecessity,aharonov2005quantum,marletto2017why-we-need}) is that they necessarily need to allow for both decoherence of quantum states and stochasticity in the classical degrees of freedom~\cite{oppenheim2021postquantum,oppenheim2022classes,galley2023consistent}, two effects that can produce observable phenomenology~\cite{oppenheim2023gravitationally}. Crucially, the decoherence and diffusion coefficients that the theory requires are not independent, but need to satisfy a relation known as the decoherence-diffusion trade-off~\cite{diosi1995quantum,oppenheim2023gravitationally}, implying that both effects cannot be made arbitrarily small. Since \emph{any} theory of fundamentally classical gravity interacting with quantum matter must satisfy the decoherence-diffusion trade-off~\cite{oppenheim2023gravitationally}, measuring its violation (by experimentally bounding both decoherence and diffusion coefficients~\cite{janse2024current}) is the simplest way to test indirectly the quantum nature of the gravitational field.

CQ theories predict that matter states decohere dynamically in the basis of interaction with the classical gravitational field, although the quantum state can remain pure, conditioned on the trajectory of the classical system~\cite{layton2024a-healthier}. In the Newtonian limit, that is simply the position basis~\cite{layton2023weak} (for massive, non-relativistic, particles), meaning that experiments in which particles are in coherent spatial superposition can be used to place strong bounds on the decoherence coefficient. 

In the next subsections, we first discuss how and when the results we have derived for the stochastic scalar field can be applied to describe the phenomenology of hybrid gravity. We then translate the upper bounds obtained from decoherence experiments, via the decoherence-diffusion trade-off, into lower bounds on the diffusion coefficient of CQ -- and analyse what this means for energy production in CQ gravity. When estimating quantities, we will move away from natural units by re-introducing factors of $G$, $c$ and $\hbar$.

\subsubsection{Validity as a model of hybrid gravity}
When considering linearised perturbation on a flat metric, i.e.:
\begin{equation}
    g_{\mu \nu}=\eta_{\mu \nu}+h_{\mu\nu} \ , \qquad |h_{\mu\nu}|\ll1 \ ,
\end{equation}
Einstein's equations in the harmonic gauge reduce to the statement that each component of the trace-subtracted metric follows a massless Klein-Gordon equation sourced by the stress-tensor of matter~\cite{wald1984general}:
\begin{equation}
\label{eq:linearised}
    \square\bar{h}_{\mu \nu} = - 16\pi G T_{\mu\nu} \ .
\end{equation}
In vacuum, the scalar mode of the metric perturbations can be gauged away, with the only physical degrees of freedom being the two polarisations of the gravitational waves. When dynamical matter is present, however, the story is more complicated -- the non-relativistic limit of the scalar mode maps to the Newtonian potential sourced by the matter distribution. For anisotropic matter, there are two independent scalar modes that are gauge-invariant -- related to the cosmological Bardeen potentials. These gauge-invariant scalar modes are not dynamical but are rather fixed by constraints equation (the Bianchi identities)~\cite{mukhanov2005physical}. The same is true for the gauge-invariant vector degrees of freedom, which are non-radiative and can in general be gauged away. Further, in cosmological backgrounds, they decay with the expansion, becoming irrelevant~\cite{mukhanov2005physical}. The only gauge-invariant mode that is indeed dynamical and follows a wave-like equation is still the transverse-traceless tensor mode -- i.e. the two degrees of freedom associated with gravitational waves~\cite{wald1984general}.

The classical-quantum path integral for general relativity has both scalar and a classical spin-2 mode~\cite{grudka2024renormalisation}, but the identification of gauge-invariant degrees of freedom in CQ theories is currently an open problem~\cite{oppenheim2022constraints}, meaning that we have not fully characterised the propagating and non-propagating modes. Our current understanding is that the tensor mode has
wave-like propagation as in GR, as does the
scalar associated with the spatial metric, while the Newtonian potential is not propagating~\cite{sajjad2025tensor}.
 
At any rate, we can consider the scalar stochastic wave equation as a toy model for Equation~\ref{eq:linearised} where we ignore the tensorial structure of the polarisation of the modes. Even if it may lack a concrete connection to CQ gravity itself, scalar fields coupled with Yukawa interactions do give rise to a $1/r$ potential, meaning that the stochastic KG field can be instructive as a simple toy model to understand the qualitative behaviour of a classical-quantum theory of gravity at least in the non-relativistic limit~\cite{carney2024classical-quantum}.  In Appendix~\ref{app:forces} we still showcase how, by treating the diffusive stochastic scalar as a toy model for the gravitational potential,
the resulting fluctuations would induce test particle to undergo Brownian motion -- and that the resulting stochastic forces would be large and measurable. 
We do not expect this to apply to the Newtonian, potential which is not  propagating, but
since there is a propagating degree of freedom which correspond to the scalar sector of the spatial metric~\cite{sajjad2025tensor}, it is important to note that we expect this force to be suppressed by $(v/c)^2$.

Moving away from the issue of constraints and physical degrees of freedom, the linearised description on flat space is guaranteed to break down as soon as the typical size of the fluctuations is larger than unity in appropriate units. As it is well-known from the theory of stochastic processes, the solution to the stochastic wave equation in 3 spatial dimensions is highly irregular, meaning that in itself the effective linear theory breaks down immediately at short scales, where non-linear terms such as those found in~\cite{oppenheim2023covariant} would therefore become important. Fortunately, non-linearities have been shown to cure the irregularity of the stochastic KG equation~\cite{oh2022three-dimensional,dalang2005hlder-sobolev}. These complexities aside, we can think of the stochastic Klein-Gordon field as providing a playground to understand some of the low-energy phenomenology of hybrid gravity theories, without much of the difficulties that arise from gauge redundancy and non-linearities that are necessary features of a complete theory.

\subsubsection{Energy density in gravitational waves}
When projected on transverse-traceless tensor modes, the evolution of the linearised metric in Equation~\ref{eq:linearised} splits into two independent stochastic KG equations for the amplitude of the two polarisations of the gravitational waves. Therefore, we can use the results from Section~\ref{sec:results} to infer the energy density of the background gravitational wave signal that CQ inevitably predicts in the weak field limit.

Indeed, these stochastic fluctuations would add up to the stochastic gravitational wave background coming from other sources~\cite{romano2017detection}. Computing the expected spectrum for the CQ signal requires taking into account an expanding background and, therefore, the induced gravitational redshift of the waves produced away from the point of observation. Such a calculation is interesting, but beyond the scope of the paper -- here we restrict to making use of the scalar results to demonstrate that, in principle, the expected signal is well within the observable domain.

Recall, we found that with a cutoff scale $\Lambda$, the average energy density produced after evolving for time $\Delta t$ is given by:
\begin{equation}
    \Delta E = \frac{Dc^5V}{8\pi G} \Lambda^3 \Delta t \ .
\end{equation}
The amplitude of the two polarisations of the tensor modes behave as two independent stochastic Klein-Gordon scalars. Moreover, the expected energy density in the CQ gravitational wave background (again, ignoring redshift due to the expansion) is given by the sum of the energy in the two independent modes:
\begin{equation}
    \rho_{GW} = \frac{2\Delta E}{V} =\frac{Dc^5}{4\pi G} \Lambda^3 \Delta t \ .
    \label{eq:EnergyGrowth}
\end{equation}
We will here take the weak field linear regime to remain valid over cosmological times, which is, as we will shortly see, a strong assumption.

Comparing this result to the observed energy density of the Universe can be used to give an order of magnitude estimate of $D \Lambda^3$ in a hybrid classical-quantum theory of gravity. In particular, the produced energy in a time $\Delta t \sim 1/H_0$ (i.e. the age of the Universe, here given in terms of the Hubble parameter $H_0$) cannot be larger than the critical energy density of the Universe $\rho_c$, for a crude bound on $D$. Taking the cutoff scale to be $\Lambda\equiv 1/\ell$, we require:
$
    \frac{H}{V}\frac{1}{\rho_c} \leq 1$ implying $ \frac{D}{\ell^3} \leq \frac{ H_0^3}{ c^3} \approx 10^{-78} \mathrm{m}^{-3}
$.

Coherence experiments on spatial superpositions currently lower bound the dimensionless diffusion coefficient\footnote{In the non-relativistic limit of local hybrid theories of gravity,  the coherence time $\tau$ of a particle in superposition of two distinct regions of volume $V$ is given  by $\tau=\frac{D_2 c^3 V}{2G_N^2N^{2/3}M^2}$ with $M$ the mass of the molecule, N the number of atoms ,$V$ the probability density over which the mass of the particle is distributed, in each disjoint region\cite{oppenheim2023gravitationally,grudka2024renormalisation}. Here, $D_2$ is taken to be dimensionless, and is one of two dimensionless constants of the theory.
Coherence measurements such as those of \cite{gerlich2011quantum} have coherence time of $\tau\approx .1\unit{s}$, using large organic molecules with total mass $M= 1.15 \times 10^{-24}kg$ and $N \sim 430$ atoms of size $r\approx 10^{-15}m$. After passing through the slits the molecule becomes delocalized in the transverse direction on the order of $d=2.7\times 10^{-7}m$. If we use $V=d^3$ we obtain $D_2\gtrapprox 10^{-71}$. This bounds differs from the one first presented in \cite{oppenheim2023gravitationally,grudka2024renormalisation} where $V\approx d^2r$ was chosen, under the unrealistic assumption that further longitudinal localisation of the molecule would not effect the interference pattern which is determined by the transverse coherence length.}
in the Newtonian potential by $D_2\gtrapprox 10^{-71}$ via the decoherence-diffusion trade-off, at the molecular length scale. As discussed in~\cite{grudka2024renormalisation}, the effective diffusion in tensor and scalar modes for a general model of CQ dynamics, can be different as the theory has two coupling constants. 
In order to get a sense of comparison, we could take $D$ to be the same order as $D_2$, which is tantamount to assuming that the diffusion $D_2$ in the Newtonian potential is of the same order as the diffusion coefficient for the tensor modes.
It is in principle allowed that the effective diffusion in scalar and tensor modes differs by several orders of magnitudes. Moreover, self-interaction will cause the diffusion coefficient to run with energy \cite{grudka2024renormalisation}, but assuming this bound holds at lower energies as well gives a rough scale at which the physics needs to change for consistency with observation. We find that, if the fluctuations gravitate, the cutoff needs to be $\ell > 10^2 \mathrm{m}$, 
which is large given that gravity has been tested to the mm scale. 

Let us stress again that this extreme lower bound on the UV cutoff for the linear model should serve only to convey the idea that these effects are roughly in the domain of observations. Indeed, it relies heavily on simplicistic assumptions, which are unlikely to hold when examining a bona fide hybrid model of gravity -- in particular that the theory is fine-tuned in such a way that the diffusion in the TT modes is of the same order of magnitude as the one in the Newtonian potential, and that the constants of the theory do not run with energy. We go over these in more detail in the Discussion section.

Another route to constrain $D$ is to compare the energy density per frequency generated by the stochasticity with direct gravitational waves observations. Penington~\cite{penington2023hipshot} has provided a useful dimensional analysis that sets a scale for possible observational tensions. He estimates the production of classical gravitational waves in hybrid theories  by noting that the amplitude of each tensor mode grows as 
\begin{equation}
    h_\omega \sim \sqrt{D \Delta t/\omega^2} \ ,
\end{equation}
which can be verified by our Equation~\ref{eq:c_kphiphi}, giving an energy density per logarithmic frequency interval
\begin{equation}
    d\rho_{\mathrm{GW}} \sim \frac{D}{G}\,\omega^3 \Delta t\, d\ln\omega \ .
\end{equation}

Using the fact that LIGO can detect characteristic strains of order $10^{-23}$ at $100$ Hz gives $D\lessapprox10^{-65}$. The bound can be improved
using  data from the first part of LIGO and VIRGO's fourth observing run, which constrains stochastic signals at $25$ Hz (\ \(\Omega_{GW}(25\mathrm{Hz})\equiv\frac{1}{\rho_C}\frac{d\rho_{\mathrm{GW}}}{d\ln\omega} \lessapprox 10^{-9}\)~\cite{abac2025upper}). Assuming that the diffusion has been active for all of our cosmic evolution, the data implies as an upper bound \(D \lesssim 10^{-68}\). This analysis has the virtue of being cutoff-independent, as the energy density per mode is of course insensitive to the physics at higher energies. 
 Combined with a lower bound \(D \gtrsim 10^{-64}\) inferred from requiring deviations from quantum mechanics to appear only below the length scales probed by the LHC (\(\sim 10^{-19}\,\mathrm{m}\)), this leaves little apparent room for viable parameter space, while using our lower bound $D\approx D_2\gtrapprox 10^{-71}$ gives significant breathing room. However, whether one can infer bounds from the LHC in this case is unclear to us, and we comment on  this in the Discussion section.

One might ask why this binned bound is much weaker than the integrated one we have discussed earlier, which led to an extreme lower bound for the UV cutoff of the linear theory with energy-independent diffusion coefficient. Indeed, $\Omega_{GW}\propto \omega^3$ is not exotic, but appears in the predicted gravitational waves background for many sources. For example, it is the expected, low-frequency, signature for the gravitational wave background generated by first-order phase transitions in the early Universe, in the frequency range that corresponds to scales that were super-horizon at the time of production~\cite{caprini2009general}. In fact, this specturm is most generally connected with uncorrelated stochastic sources of gravitational waves in post-inflationary evolution~\cite{caprini2018cosmological}. 

Crucially, however, this behaviour naturally caps off at the typical correlation scale of the system $f^*$, whatever that may be, and transitions at higher energies to a spectrum whose integral converges -- the details depend on the specific source considered~\cite{caprini2018cosmological}. Therefore, in standard analysis, the integrated bounds coming from physical extended sources are never drastic with respect to the ones derived from observations in a specific frequency range. In the linear regime of the stochastic model we study, however, the naive frequency spectrum grows cubically for arbitrarily high energies -- until new physics comes in.

\section{Discussion} 
\label{sec:discussion}
Classical-quantum theories of gravity -- in which a classical metric interacts with quantum fields -- predict stochastic fluctuations in the gravitational field. A simple model to study the phenomenology of CQ theories is two scalar KG field (one classical and one quantum) interacting with Yukawa coupling. In this article, we have focused on the classical (free sector) of this theory -- the stochastic Klein-Gordon equation.

We began with a review of the classical Klein-Gordon thermal field, and the damped stochastic Klein-Gordon equation, for which the thermal state is a fixed point. We then removed friction, and discussed the non-dissipative system -- which corresponds to the classical sector of the CQ scalar Yukawa theory. We first computed the two-point function of the system and showed it has a $1/r$ scaling for spacelike separated points -- with the intensity of the covariance increasing linearly with the time elapsed from the initial slice. We similarly showed that the energy of the field grows linearly with time -- though with an infinite rate if no UV cutoff exists. This is related to the infinite energy of the field itself, due to the contact divergence. A similar contact divergence occurs for quantum fields -- it is what gives rise to the cosmological constant problem.  We finally computed the exact solution mode-by-mode for both the massive and massless classical field. We concluded by showing that the fluctuations have measurable physical effects. Indeed, we calculated the expected energy density of the background gravitational wave signal predicted by CQ theories, showing that for it to be consistent with experiments, either the linear approximation needs to break down at macroscopic scale, or the parameters of the theory need to have values such that the induced diffusion coefficient for the scalar mode is much larger than the one for the tensor modes.

In the discussion by Penington~\cite{penington2023hipshot} that we discussed when analysing the compatibility of CQ with observations, he combined the upper bound of \(D \lesssim 10^{-68}\) obtained through LIGO data with a lower bound of \(D \gtrsim 10^{-64}\), inferred from requiring deviations from quantum mechanics to appear only below the length scales probed by the LHC (\(\sim 10^{-19}\,\mathrm{m}\)). This leaves little apparent room for viable parameter space. By contrast, our analysis draws its lower bound from laboratory decoherence experiments, which currently constrain the diffusion parameter in the Newtonian sector to \(D_2 \gtrsim 10^{-71}\), and an anticipation that the diffusion coefficient $D$ for tensor modes will be of the same order as that for scalar modes \(D_2\approx D\) -- giving a few orders of magnitude of headroom. We view decoherence experiments as providing a reasonable bound, because the short distance behaviour of the theory is not well understood. Here, we expect anomalous heating bounds in the quantum degrees of freedoms~\cite{donadi2021underground} to give the most stringent bound. It is unclear  to what extent measurements from the LHC, or from considerations of anomalous heating, directly constrain $D$ (the coupling to tensor modes). This is because the decoherence-diffusion trade-off for tensor modes implies that $G_N^2/D$  gives the strength of decoherence in the strain components of the stress-energy tensor e.g. via $\hat{T}_{ij}\hat{T}^{ij}$, which suggests that experiments which are sensitive to strain noise could be more relevant than collider physics. While the LHC is likely to find novel forces, it's not clear whether it would discover decoherence channels, especially one which causes decoherence due to emission of gravitational waves -- as it is the case here and as we would expect to be the case in quantum gravity when tracing out the graviton.
In this context, it is worth noting that the Planck length is not necessarily the relevant length scale of the theory, which is instead set by the coupling constants $G_N^2/D_2$ and the corresponding one for tensor modes~\cite{oppenheim2023covariant}.

Anomalous heating bounds will also require a greater understanding of the short distance behaviour of the classical-quantum theory, and in particular, the non-linear and classical-quantum interplay which will dominate at short distances. We especially need to understand the length scale that is set in the theory by dimensional transmutation~\cite{grudka2024renormalisation}. 
As shown, stronger upper bounds come from the integrated energy-density \(\rho_{\mathrm{GW}}\!\propto\! D\,\Lambda^3 T\). However, these rely on a number of simplifications, and these bounds should only be interpreted as a guide. We have here assumed $D\approx D_2$, ignored the classical-quantum coupling, and assumed we are in Minkowski space. 
Including Hubble friction and the dilution of radiation with expansion, will suppress the energy density relative to the naive flat-space estimate of Equation~\eqref{eq:EnergyGrowth}. However, it is likely that a complete treatment -- tracking stochastic wave production on an expanding background and accounting for non-linear couplings between the classical and quantum sectors -- will be needed to assess observational viability. 
While the theory must ultimately satisfy stringent bounds, its cosmological behaviour has the potential to admit a self-consistent window compatible with experiment. This reinforces the idea that cosmological observations combined with tabletop experiments, provide a strong probe into the quantum nature of spacetime.

This work highlights several possible interesting directions to pursue. A natural one is to include the quantum field in the linear regime -- a CQ model of classical spin-2 fields interacting with a quantum scalar field, for example. Studying such a system is a crucial step in the formulation of a fundamental CQ theory of gravity, and a central aim would be to  place a lower bound on the diffusion coefficients due to anomalous strains in the quantum sector. Together with the upper bounds on diffusion coming from cosmology, this could be used to rule out a fundamentally classical spacetime. Here, the interplay between the quantum and classical fields at short distance are expected to be crucial to understand the UV theory.

Such a model also has applications to the theory of cosmological perturbations, with the derivation of CQ predictions for the spectrum of the cosmic microwave background and gravitational wave background being key objectives. These would provide formidable tests for hybrid theories of gravity -- and allow for a more thorough investigation of the proposal that stochastic fluctuations in the gravitational field can act as cold dark matter~\cite{oppenheim2024emergence,oppenheim2024anomalous}. Coming up with independent tests on the diffusion in the Newtonian potential and in the gravitational wave background can also allow us to fix the free CQ parameter $\beta$~\cite{grudka2024renormalisation}, which, as discussed, can amplify the noise in one sector with respect to the other. Positivity constraints force $\beta\leq 1/3$. If tests on the diffusion coefficient of scalar and tensor modes suggest $\beta$ needs to violate this bound to be in agreement with experiments, the theory would be ruled out.

Another natural direction is to study the renormalisation properties of the CQ scalar Yukawa theory. One path-integral formulation of CQ gravity  is expected to be free of the renormalisation issues that plague Einstein's gravity due to the analogies with the quadratic gravity action\cite{grudka2024renormalisation}, whose quantum theory is known to be renormalisable~\cite{stelle1977renormalization}. A virtue of the CQ version is that it is free of the ghosts that plague the quadratic quantum theory. Nonetheless, no explicit renormalisation analysis has ever been performed on a CQ field theory -- the scalar Yukawa model seems the ideal playground to approach the problem by building on known results for classical statistical field theory and open QFT~\cite{fisher1998renormalization,baidya2017renormalization}. A fundamental question is whether the renormalisation group can preserve the decoherence-diffusion trade-off, the crucial consistency condition of CQ theories. Moreover, as we discuss at length in Appendix~\ref{app:forces} when estimating the induced forces from the classical stochastic fluctuations on extended object, it is important to address whether self-interaction vertices (either from non-linearities in the classical equations or induced from quantum backreaction) can indeed curb the irregularity of the free stochastic wave equation -- building on some existing formal results~\cite{conus2008the-non-linear,oh2022three-dimensional}. It is also crucial to understand the relationship between the Newtonian potential, and any propagating modes.

Finally, CQ theories have been shown to emerge as effective theories when the partial classical limit of a bipartite quantum system~\cite{layton2024classical,layton2025effective} is taken. Extending this result to the case of field theories is crucial to understand better the regime of validity of the mean-field Einstein's equations~\cite{ahmed2024semiclassical,kuo1993semiclassical}, and of CQ models as \emph{effective} theories in the limit that the gravitational field behaves classically. This is in the spirit of including both decoherence and diffusion effects, going beyond the stochastic corrections to the mean-field semiclassical Eisntein's equations that are considered in the formalism of stochastic gravity~\cite{hu2008stochastic}.

\vspace{3mm} 

\noindent\section*{Acknowledgements}
The authors would like to thank Daniel Carney, Luke Caldwell,  Rhys Evans, Stephan Hogan, Alex Jenkins,  Johannes Noller, Geoff Penington, Andrew Pontzen, Muhammad Sajjad, and Gautam Satischandram, for insightful discussions.
\newpage

\appendix

\section{CQ basics}
\label{app:CQ_basics}
CQ theories describe the evolution of a collection of $2N$ classical degrees of freedom $z^i \in \mathcal{M}$ -- where $\mathcal{M}$ is the $2N$-dimensional phase-space -- with a quantum system whose reduced state and observables can be described in terms of a Hilbert space $\mathcal{H}$. The fundamental object of the hybrid formalism is the CQ state, a phase-space-valued super-density operator $\varrho$ obeying the following normalisation conditions:
\begin{equation}
    \text{Tr}(\varrho) = p(z) \in \mathcal{P}(\mathcal{M}) \  , \qquad \int_{\mathcal{M}}\varrho(z)dz = \rho \in \mathcal{D_2}(\mathcal{H}) \ ,
\end{equation}
where $\mathcal{P}(\mathcal{M})$ is the space of normalised non-negative densities (i.e. well-defined probability distributions) over phase space, whilst $\mathcal{D_2}(\mathcal{H})$ is the set of bounded trace-one positive semi-definite linear operators over $\mathcal{H}$ -- namely density operators. This implies the obvious normalisation condition:
\begin{equation}
   \int_{\mathcal{M}} \text{Tr}(\varrho) dz = 1 \ .
\end{equation}
In order to respect the statistical interpretation of the state, consistent CQ evolution needs to be linear, trace-preserving and completely-positive. This guarantees that CQ states are mapped to other valid CQ states by the evolution operator. A further condition, the one of a Markovian (i.e. time-local) dynamics, is often assumed to simplify further the form of the evolution equations. In fact, it can be proved that under these four conditions the form of the dynamics is unique~\cite{oppenheim2022classes}. In particular, the evolution of a hybrid CQ state $\varrho$ for a collection of classical fields $\phi^i$ interacting with quantum ones $\psi^a$ can be expressed in the path integral formalism as~\cite{oppenheim2023path,oppenheim2023covariant}:
\begin{equation}
\label{eq:cq_pi}    \varrho(\phi_f,\psi_f,t_t) = \int\mathcal{D}\psi_L \mathcal{D}\psi_R \mathcal{D}\phi \mathcal{D}\tilde{\phi} e^{I_{CQ}[\psi_L,\psi_R,\phi,\tilde{\phi}]}  \varrho(\phi_0,\psi_0,t_0) \ .
\end{equation}
Here, $L/R$ subscript indicate the left and right branches of the path integral in the quantum sector (in Schwinger-Keldysh sense), whilst $\tilde{\phi}$ are a collection of purely imaginary-valued auxiliary fields known in the theory of stochastic processes as \textit{response variables}. These are not physical dynamical fields, but are there to encode the interaction of the fundamental degree of freedom $\phi$ with the noise field. The classical-quantum action to propagate the state from an initial hypersurface $\Sigma_0$ to a final one $\Sigma_f$ is given by:
\begin{equation}
\begin{split}
\label{eq:CQ_act}
    I_{CQ}[\psi_L,\psi_R,\phi,\tilde{\phi}] = \int_{\Sigma_0}^{\Sigma_f} d^4x \left[ i\left(\mathcal{L}_0^{Q}[\psi_L] + \mathcal{L}_{int}[\psi_L,\phi]-\mathcal{L}_0^{Q}[\psi_R]-\mathcal{L}_{int}[\psi_R,\phi] \right)\right. \\
    -\left. \frac{D_0}{2}\left(\frac{\delta \mathcal{L}_{int}[\psi_L,\phi]}{\delta \phi}-\frac{\delta \mathcal{L}_{int}[\psi_R,\phi]}{\delta \phi}  \right)^2 \right. \\
    \left. +\frac{D_2}{2}  \tilde{z}^2 -\tilde{z}\left(\frac{\delta \mathcal{L}^{C}[\phi]}{\delta \phi}+\frac{1}{2}\frac{\delta \mathcal{L}_{int}[\psi_L,\phi]}{\delta \phi}+\frac{1}{2}\frac{\delta \mathcal{L}_{int}[\psi_R,\phi]}{\delta \phi} \right)\right] \ ,
\end{split}
\end{equation}
where the classical stochastic sector is represented in the Martin-Siggia-Rose (MSR) formalism~\cite{martin1973statistical,hertz2016path} -- and we have assumed that we can integrate out both classical and quantum momenta. One can easily integrate over the quadratic response field, and obtain the Onsager-Machlup form of the CQ path integral, as in~\cite{oppenheim2023covariant,oppenheim2023path}. However, the virtue of doubling the classical degrees of freedom is that it establishes a clear correspondence between the quantum and classical sectors -- and makes the action more apt for perturbation theory. Here, $D_2$ and $D_0$ are the respectively the diffusion and decoherence coefficients of the theory (we have taken both the diffusion and decoherence kernels to be $\delta$-function in spacetime to comply with Lorentz invariance). Indeed, the first line corresponds with the unitary part of the quantum evolution, whilst the second line represents the action of a decoherence channel -- since it suppresses off-diagonal components of the density matrix ($\psi_L \neq \psi_R$). The final line describes the dynamics of the classical state (including the backreaction due to the quantum system), as one can identify it simply as the MSR action of a stochastic classical system -- with the quantum one acting as a source. 

Mixing quantum mechanics and stochastic processes is a practice done for a variety of reasons. Other than the obvious relation between the Schr\"odinger equation in imaginary time and the Fokker Planck equation~\cite{nelson1966derivation,nagasawa1993schradinger}, there is no shortness of attempts to reproduce the phenomenology of quantum mechanics by using underlying classical stochastic processes. For example, stochastic quantisation~\cite{parisi1981perturbation} uses diffusion in a fictitious extra time dimension to prepare the two-point functions of quantum fields from classical fields, although a similar result can be obtained from averaging over random phases of an ensemble of classical deterministic fields~\cite{holdom2006approaching}. Moreover, the parallelism between classical stochastic processes and quantum mechanics can also be used the other way around -- i.e. borrowing techniques from quantum theory to solve problems in probability theory~\cite{baez2019quantum}. It is however important to stress that, in the CQ framework, the classical variable represents the state of a physical, observable, classical system, and that the diffusion happens in real time.

Crucially, $D_0$ and $D_2$ are not independent. In order for the dynamics to be linear, trace-preserving and completely positive, they need to satisfy the so-called decoherence-diffusion trade-off -- a relation between the size of the stochasticity in the classical system and the typical decoherence rate in the quantum sector. The details are not important for the present paper, and we refer to~\cite{oppenheim2022classes,oppenheim2023path} for the interested reader. However, the diffusion and decoherence matrices need to satisfy: 
\begin{equation}
\label{eq:deco-diff}
    D_2  D_0\succeq 1 \ ,
\end{equation}
meaning that both effects cannot be made small. Interestingly, when the trade-off is saturated -- i.e. the left hand side of Equation~\ref{eq:deco-diff} \emph{equals} the right hand side -- the hybrid dynamics does not destroy information~\cite{layton2024a-healthier}. By keeping a record of the state of the classical system, the quantum state remains pure.

In the main body of the paper, we simply call the diffusion coefficient $D$, rather than $D_2$. This is not only to streamline notation, but also to sever any connection between the fundamental $D_2$ in a CQ theory of gravity and the effective diffusion coefficient in any particular mode of linearised gravity. For example, the scalar and the tensor modes can in principle have significantly different effective diffusion coefficients for the same $D_2$, due to an additional free parameter of CQ gravity, which we call $\beta$, that can amplify noise in one mode with respect to the other~\cite{grudka2024renormalisation}.

\subsection{CQ scalar Yukawa}
The stochastic Klein-Gordon field we study in the main body is of interest because it is one of the building block for the simplest CQ field theory, the hybrid Yukawa model of~\cite{grudka2024renormalisation} c.f.~\cite{oppenheim2023covariant}. Scattering in this model was recently discussed in~\cite{carney2024classical-quantum} where it was shown that stochastic fluctuations can affect non-trivially scattering probabilities in CQ theories. The CQ scalar Yukawa theory corresponds to two dynamical Klein-Gordon scalars, the classical $\phi$ and quantum $\psi$. Specifically, the CQ proto-action can be obtained with the following choices:
\begin{equation}
    \mathcal{L}^Q_0[\psi] = \frac{1}{2} \psi (\square +m^2 ) \psi  \ , \qquad \mathcal{L}^C_0[\phi]= \frac{1}{2} \phi (\square + m^2 ) \phi
\end{equation}
and
\begin{equation}
    \mathcal{L}_{int}[\psi,\phi] = \lambda \phi \psi^2
\end{equation}

In the main body, we ignore completely the quantum sector, and focus on the free part of the classical sector of the theory, whose understanding is crucial to construct perturbation theory and explore the renormalisation of the model. Moreover, as we discuss in Subsection~\ref{sec:pheno}, it encodes the background fluctuations in the traceless-tensor modes of the linearised metric generated in CQ by the fundamental noise required by the decoherence-diffusion trade-off.

\section{Brownian motion}
\label{app:brownian}
\subsection*{SDEs perspective}
As an illustrative example of the results presented here, consider a particle undergoing Brownian motion:
\begin{equation}
\label{eq:brownian}
    \ddot{q} = \xi(t) \ ,
\end{equation}
where:
\begin{equation}
    \mathbb{E}[\xi(t)]=0 \ , \qquad \mathbb{E}[\xi(t) \xi(t')] = D \delta(t-t') \ .
\end{equation}
We initialise the state at $t=t_0$ such that $q(t_0)=0$ and $\dot{q}(t_0)=0$. Essentially:
\begin{equation}
    P(q,\dot{q},t_0) = \delta(q)\delta(\dot{q}) \ .
\end{equation}
We then let the system evolve following Equation~\ref{eq:brownian} up to some future time $t$. The covariance of the stochastic process is then defined as:
\begin{equation}
    C(t,s) = \mathbb{E}[q(t),q(s)|q(t_0)=\dot{q}(t_0)=0] \ .
\end{equation}
Given a realisation of the stochastic field $\xi$, we can reconstruct the trajectory of the Brownian particle by simply convoluting with the retarded Green's function of the equation of motion:
\begin{equation}
    \label{eq:ret_green}
     q(t) = \int d t' G_R(t,t') \xi(t') \theta(t'-t_0) \ ,
\end{equation}
where the theta function is there to impose the BC's and:
\begin{equation}
    G_R(t,t') = \theta(t-t') (t-t') \ .
\end{equation}

Therefore, given the boundary conditions, we simply have (assume $t\geq s$):
\begin{align}
    C(t,s|t_0) &= \int_{t_0}^\infty d\tau \int_{t_0}^\infty  d\tau' G_R(t,\tau) G_R(s,\tau') \mathbf{E}[\xi(\tau)\xi(\tau')] \\
    \label{eq:convolution}
    & =D \int_{t_0}^\infty d\tau \theta(t-\tau) \theta(s-\tau) (t-\tau) (s-\tau)
\end{align}
Since $t\geq s$, the first $\theta$-function is irrelevant, whilst the second one sets the upper integration limit:
\begin{align}
   C(t,s|t_0) &= D\int_{t_0}^{s} d\tau \ G_R(t,\tau) G_R(s,\tau) \\
   & = D \int_{t_0}^{s} d\tau (t-\tau) (s-\tau) \\
   & = \frac{D}{6}(s-t_0)^2(3t-s-2t_0) \ ,
\end{align}
or, in explicit powers of $s$:
\begin{equation}
    C(t,s) = \frac{D}{6} [-s^3+3 t s^2 + 3st_0(t_0-2t) +t_0^2(3t-2t_0)] \ .
\end{equation}
The variance, instead:
\begin{equation}
V(t;t_0) \equiv C(t,t;t_0) = \frac{D}{3} (t^3-3 t_0 t^2+3  t_0^2 t-t_0^3) \ .
\end{equation}

Clearly, as $t_0 \to -\infty$ both variance and covariance diverge.

\subsubsection*{Fourier representation of the propagator}
We know derive the same result by performing the convolution in Fourier domain, uncovering the pole prescription of the propagator in the complex plane. Let's begin with the Fourier representation of the retarded propagator. This is given by:
\begin{equation}
    G_R(t-s) = \int\frac{d \omega}{2\pi} \mathcal{G}_R(\omega) e^{-i\omega(t-s)} \ ,
\end{equation}
where:
\begin{equation}
    \mathcal{G}_R(\omega) = \frac{-1}{(\omega+i\epsilon)^2} \ .
\end{equation}
To see this, first note that for $t < s$, the Fourier integral vanishes since the complex contour of integration (closing from above, to make the contribution from the semicircle limit to zero). For $t>s$, however, the contour encloses the $\omega=-i\epsilon$ pole of order $p=2$. Using the residue theorem we see:
\begin{align}
    \int\frac{d \omega}{2\pi} \frac{-1}{(\omega+i\epsilon)^2} e^{-i\omega(t-s)} &= -i \text{Res}\left[\frac{1}{(\omega+i\epsilon)^2} e^{-i\omega(t-s)},-i\epsilon\right] \\
    & = t-s
\end{align}
Combining the two time-ordered result we indeed obtain Equation~\ref{eq:ret_green}.

Now, let's derive the Fourier representation of $C(t,s)$. To do this, we start from Equation~\ref{eq:convolution} and substitute the Fourier representation of the retarded propagators:
\begin{align}
    C(t,s|t_0)&=D \int_{t_0}^\infty d\tau \int_{-\infty}^{\infty}\frac{d\omega}{2\pi}\int_{-\infty}^{\infty}\frac{d\omega'}{2\pi} \frac{1}{(\omega+i\epsilon)^2}
    \frac{1}{(\omega'+i\epsilon)^2}e^{-i\omega(t-\tau)} e^{-i\omega'(s-\tau)} \\
    & = D \int_{-\infty}^{\infty}\frac{d\omega}{2\pi}\int_{-\infty}^{\infty}\frac{d\omega'}{2\pi} \frac{1}{(\omega+i\epsilon)^2}
    \frac{1}{(\omega'+i\epsilon)^2}
    e^{-i\omega t}
    e^{-i\omega's}
    \int_{t_0}^\infty  d\tau e^{i(\omega+\omega')\tau} \\
    \label{eq:conv_finite}
    & = D \int_{-\infty}^{\infty}\frac{d\omega}{2\pi}\int_{-\infty}^{\infty}\frac{d\omega'}{2\pi} \frac{1}{(\omega+i\epsilon)^2}
    \frac{1}{(\omega'+i\epsilon)^2}
    e^{-i\omega t}
    e^{-i\omega's}
    \int_{0}^\infty  d\tau e^{i(\omega+\omega')(\tau+t_0)} \\
    & = D \int_{-\infty}^{\infty}\frac{d\omega}{2\pi}\int_{-\infty}^{\infty}\frac{d\omega'}{2\pi} \frac{1}{(\omega+i\epsilon)^2}
    \frac{1}{(\omega'+i\epsilon)^2}
    e^{-i\omega (t-t_0)}
    e^{-i\omega'(s-t_0)}
    \left(\pi \delta(\omega+\omega')+ \mathbf{P}\frac{i}{\omega+\omega'} \right) \\
    & = \frac{1}{2}C_\infty(t,s) + \Delta C(t,s|t_0)
\end{align}
Here we have split the integral in the infinite time (the integral involving the delta function kills all the $t_0$ dependence, and is exactly half of the integral resulting from the $t_0 \to -\infty$ limit) and finite time effects. Note that here $\mathbf{P}$ indicates that the principal value of the integral needs to be extracted for such a pole.

Let's begin with:
\begin{align}
    C_\infty(t,s) 
    &=  D \int_{-\infty}^{\infty}\frac{d\omega}{2\pi}\int_{-\infty}^{\infty}\frac{d\omega'}{2\pi} \frac{1}{(\omega+i\epsilon)^2}
    \frac{1}{(\omega'+i\epsilon)^2}
    e^{-i\omega (t-t_0)}
    e^{-i\omega'(s-t_0)}
    2\pi \delta(\omega+\omega') \\
    &=D \int_{-\infty}^{\infty}\frac{d\omega}{2\pi} \frac{1}{(\omega+i\epsilon)^2}
    \frac{1}{(\omega-i\epsilon)^2}
    e^{-i\omega (t-s)} \\
    & = \int_{-\infty}^{\infty}\frac{d\omega}{2\pi} \mathcal{C}_{\infty} (\omega)
    e^{-i\omega (t-s)}
\end{align}
where:
\begin{equation}
    \mathcal{C}_{\infty} (\omega) = \frac{D}{(\omega+i\epsilon)^2 (\omega-i\epsilon)^2} \ .
\end{equation}
Note, this is different from the Feynman mod-squared (FM2) prescription proposed in~\cite{grudka2024renormalisation}. In fact, the latter is given by:
\begin{equation}
    \mathcal{C}_\infty^{\text{mod2}} (\omega) = \frac{D}{(\omega^2+i\epsilon) (\omega^2-i\epsilon)} \neq \mathcal{C}_{\infty} (\omega) \ .
\end{equation}
The biggest difference between the two prescriptions is the nature of the poles. The 2-point function obtained as convolution of 2 retarded Green's function (RM2) has two second order poles, whilst the FM2 prescription involves four simple poles.

The finite time effects, instead, are given by:
\begin{equation}
\begin{split}
    \Delta C(t,s|t_0) &= \mathbf{P}  \int_{-\infty}^{\infty}\frac{d\omega}{2\pi}\int_{-\infty}^{\infty}\frac{d\omega'}{2\pi} \frac{i D}{(\omega+i\epsilon)^2(\omega'+i\epsilon)^2(\omega+\omega')}
    e^{-i\omega (t-t_0)}
    e^{-i\omega'(s-t_0)} \\
    & = \frac{1}{2} \int_{-\infty}^{\infty}\frac{d\omega}{2\pi} \mathcal{C}_\infty (\omega)  e^{-i\omega(t-s)} \\
    & \qquad- D e^{-\epsilon(s-t_0)}\int_{-\infty}^{\infty}\frac{d\omega}{2\pi}\left(i (s-t_0) + \frac{1}{\omega-i\epsilon} \right) \frac{e^{-i\omega(t-t_0)}}{(\omega + i \epsilon)^2 (\omega-i\epsilon)} 
\end{split}
\end{equation}
\subsection*{Performing the inverse Fourier transform}
Let's begin with the infinite time effects. For $t\geq s$ close below, picking up the double $\omega = - i \epsilon$ pole. Using the residue theorem we get:
\begin{equation}
    C_\infty(t,s) = \lim_{\epsilon \to0}\frac{e^{-\epsilon(t-s)}}{4 \epsilon^2}\left(t-s+\frac{1}{\epsilon} \right) \ .
\end{equation}
On the other hand:
\begin{equation}
\begin{split}
    D e^{-\epsilon(s-t_0)}\int_{-\infty}^{\infty}\frac{d\omega}{2\pi}\left(i (s-t_0) + \frac{1}{\omega-i\epsilon} \right) \frac{1}{(\omega + i \epsilon)^2 (\omega-i\epsilon)} e^{-i\omega(t-t_0)} = \\
    \lim_{\epsilon \to0} \frac{D e^{-\epsilon(t+s-2t_0)}}{2\epsilon}\left[(s-t_0) \left(t-t_0+\frac{1}{2\epsilon}\right) + \frac{1}{2\epsilon}\left(t-t_0+\frac{1}{\epsilon} \right)\right] \ .
\end{split}
\end{equation}
Plugging all together and keeping terms up to $\mathcal{O}(\epsilon^0)$ we get:
\begin{equation}
    C(t,s|t_0) = \frac{D}{6}(s-t_0)^2(3t-s-2t_0) \ .
\end{equation}
This is indeed the correct solution, and no divergent term survives. Recall, once more, that here $t\geq s$ was assumed.

\section{Integrals in Fourier space}
\label{app:four_int}
We assume the field (and its conjugate momentum) is initialised on an initial slice $t_0$ to vanish everywhere, i.e. $\phi(\underline{x},t_0)=\dot{\phi}(\underline{x},t_0)=0$. Then the covariance is given by:
\begin{equation}
    \mathcal{C}(x,t;t_0) = D \int_{t_0}^\infty dz_0 \int d^3z \int \frac{d^4p}{(2\pi)^4}\int \frac{d^4k}{(2\pi)^4} \frac{e^{-ip(x-z)}e^{-ik(y-z)}}{\left[(p_0+i\epsilon)^2-E(\underline{p})^2 \right]\left[(k_0+i\epsilon)^2-E(\underline{k})^2 \right]} \ .
\end{equation}
Now, let's focus on the $z_0$ integration first:
\begin{equation}
    \int_{t_0}^\infty dz_0 e^{iz_0(p_0+k_0)} = e^{i(p_0+k_0)t_0}\int_0^\infty d\tau e^{i\tau(p_0+k_0)} = e^{i(p_0+k_0)t_0} \left(\pi \delta(p_0+k_0)+ i \frac{\mathbf{P}}{k_0+p_0} \right) ,
\end{equation}
where the capital $\mathbf{P}$ stands for the \textit{principal value} of the integral for the corresponding pole.  The spatial $z$ integral gives:
\begin{equation}
    \int d^3z e^{-iz(p+k)} = (2\pi)^3 \delta(\underline{p}+\underline{k}).
\end{equation}
Combining:
\begin{equation}
\mathcal{C}(x,t;t_0) = \frac{1}{2}\mathcal{C}_\infty + \bar{\mathcal{C}},
\end{equation}
where:
\begin{equation}
\begin{split}
    \mathcal{C}_\infty&= D \int \frac{d^4p}{(2\pi)^4}\int \frac{d^4k}{(2\pi)^4} \frac{e^{-ipx}e^{-iky}}{\left[(p_0+i\epsilon)^2-E(\underline{p})^2 \right]\left[(k_0+i\epsilon)^2-E(\underline{k})^2 \right]} (2\pi)^4 \delta(p+k) \\
    &= D \int \frac{d^4p}{(2\pi)^4} \frac{e^{-ip(x-y)}}{\left[(p_0+i\epsilon)^2-E(\underline{p})^2 \right]\left[(p_0-i\epsilon)^2-E(\underline{p})^2 \right]}  \ ,
\end{split}
\end{equation}
whilst:
\begin{equation}
\begin{split}
    \bar{\mathcal{C}} &= i D \int \frac{d^4p}{(2\pi)^4}\int \frac{d^4k}{(2\pi)^4} \frac{e^{-ipx}e^{-iky}}{\left[(p_0+i\epsilon)^2-E(\underline{p})^2 \right]\left[(k_0+i\epsilon)^2-E(\underline{k})^2 \right]} (2\pi)^3 \delta(\underline{p}+\underline{k}) \frac{\mathbf{P}}{p_0+k_0}\\
    &= i D \int \frac{d^4p}{(2\pi)^4}\int \frac{dk_0}{2\pi} \frac{ e^{i\underline{p}(\underline{x}-\underline{y})}e^{-ip_0x}e^{-ik_0y}}{\left[(p_0+i\epsilon)^2-E(\underline{p})^2 \right]\left[(k_0+i\epsilon)^2-E(\underline{p})^2 \right]} \frac{\mathbf{P}}{p_0+k_0}
\end{split}
\end{equation}
\subsection*{Infinite-time terms}
Let's first compute $\mathcal{C}_\infty$ for the massive KG field corresponding to the retarded mod-squared prescription:
\begin{equation}
\begin{split}
    \mathcal{C}_{\infty}(x,y) &= D\int \frac{d^4 p}{(2\pi)^4}\frac{e^{-ip(x-y)}}{[(p_0-i\epsilon)^2-E(p)^2][(p_0+i\epsilon)^2-E(p)^2]} \\
    &=D \int \frac{d^4 p}{(2\pi)^4}\frac{e^{-ip(x-y)}}{(p_0+ i\epsilon + E(\underline{p}))(p_0+ i\epsilon - E(\underline{p}))(p_0- i\epsilon + E(\underline{p}))(p_0- i\epsilon - E(\underline{p}))}
\end{split}
\end{equation}
Throughout we assume the time ordering of the events as $x_0 \geq y_0$. This means that we close the complex contour from below (clockwise) picking up contributions from the poles in the lower-half plane. First, let's integrate over energy:
\begin{equation}
\begin{split}
    & \int_{-\infty}^{+\infty} \frac{d p_0}{2\pi} \frac{e^{-ip_0(x^0-y^0)}}{(p_0+ i\epsilon + E(\underline{p}))(p_0+ i\epsilon - E(\underline{p}))(p_0- i\epsilon + E(\underline{p}))(p_0- i\epsilon - E(\underline{p}))} \\
    & = -i\left[\text{Res}\left(f(p_0) ,p_0 =E(\underline{p})-i\epsilon \right) + \text{Res}\left(f(p_0) ,p_0 = -E(\underline{p})-i\epsilon \right) \right] \\
    & = \frac{-i}{-8i \epsilon E(\underline{p})} e^{-\epsilon(x^0-y^0)}\left[ \frac{e^{-iE(\underline{p})(x^0-y^0)}}{E(\underline{p})-i\epsilon}+\frac{e^{-iE(\underline{p})(x^0-y^0)}}{E(\underline{p})+i\epsilon}\right] \\
    & = \frac{1}{4\epsilon E(\underline{p})(E(\underline{p})^2+\epsilon^2)} e^{-\epsilon(x^0-y^0)} \left[E(\underline{p})\cos\left(E(\underline{p}) (x^0-y^0) \right)+\epsilon \sin \left(E(\underline{p})(x^0-y^0) \right) \right] \ .
\end{split}
\end{equation}
Moreover, we can integrate the angular coordinates of the 3-momentum:
\begin{equation}
    \int_0^{2\pi}d\phi \int_0^{\pi}d\theta \sin\theta \ e^{i\underline{p}(\underline{x}-\underline{y})} = \int_0^{2\pi}d\phi \int_0^{\pi}d\theta \sin\theta \  e^{ip|\underline{x}-\underline{y}|\cos\theta} = 4\pi \frac{\sin(p|\underline{x}-\underline{y}|)}{p|\underline{x}-\underline{y}|} .
\end{equation}
Finally, recall the following limit:
\begin{equation}
    \lim_{\epsilon\to 0} \frac{\epsilon}{x^2+\epsilon^2} = \pi \delta(x) \ .
\end{equation}

We can now combine these results to obtain:
\begin{equation}
\begin{split}
    \mathcal{C}_\infty &= \frac{D}{8\pi^2} \frac{e^{-\epsilon(x^0-y^0)}}{|\underline{x}-\underline{y}|} \left(\frac{1}{\epsilon}I_1 + I_2\right) \\
    &= \frac{D}{8\pi^2} \frac{1}{|\underline{x}-\underline{y}|} \left[\left(\frac{1}{\epsilon}- (x^0-y^0) \right)I_1 + I_2 +\pi \frac{(x^0-y^0)^2}{2}I_{\delta1} - \pi(x^0-y^0) I_{\delta 2} \right] + \mathcal{O}(\epsilon)
\end{split}
\end{equation} 
Here we have defined:
\begin{equation}
\label{eq:I1}
    I_1 = \int_0^\infty dp \frac{p}{E(p)^2} \cos\left(E(p) (x^0-y^0) \right) \sin(p|\underline{x}-\underline{y}|),
\end{equation}

\begin{equation}
    I_2 = \int_0^\infty dp \frac{p}{E(p)^3} \sin\left(E(p) (x^0-y^0) \right) \sin(p|\underline{x}-\underline{y}|)
\end{equation}

\begin{equation}
    I_{\delta 1} = \int_0^\infty dp \ \delta(E(p)) \ p \cos\left(E(p) (x^0-y^0) \right) \sin(p|\underline{x}-\underline{y}|)=0,
\end{equation}

\begin{equation}
    I_{\delta2} = \int_0^\infty  dp \  \delta(E(p)) \ \frac{p}{E(p)} \sin\left(E(p) (x^0-y^0) \right) \sin(p|\underline{x}-\underline{y}|)=0 \ .
\end{equation}
Note that both $ I_{\delta 1}= I_{\delta 2}=0$ for any $m \geq0$, since for the massive case the delta function condition is never satisfied, whilst for the $m=0$ case at $p=0$ the integrand is vanishing. Therefore, in the $\epsilon\to 0$ limit, the infinite-time component of the covariance is given by:
\begin{equation}
    \mathcal{C}_\infty = \frac{D}{8\pi^2} \frac{1}{|\underline{x}-\underline{y}|} \left[\left(\frac{1}{\epsilon}- (x^0-y^0) \right)I_1 + I_2 \right] \ .
\end{equation}
We were not able to find a closed analytical solution for both $I_1$ and $I_2$ for arbitrary mass.

\subsection*{Finite-time contribution}
Let's now compute the finite-time contributions to the two-point function. This corresponds to evaluating the following integral:
\begin{equation}
    \tilde{\mathcal{C}} = i D \int d^3z \int \frac{d^4p}{(2\pi)^4}\int \frac{d^4k}{(2\pi)^4} \frac{e^{-ipx}e^{-iky} e^{i(p_0+k_0)t_0} e^{-i(\underline{p}+\underline{k}).\underline{z}}}{\left[(p_0+i\epsilon)^2-E(\underline{p})^2 \right]\left[(k_0+i\epsilon)^2-E(\underline{k})^2 \right]} \frac{\mathbf{P}}{k_0 + p_0} \ .
\end{equation}
Performing the integral and extracting the principal value one gets:
\begin{equation}
    \tilde{\mathcal{C}} = \frac{1}{2}\mathcal{C}_\infty + \Delta\mathcal{C} \ ,
\end{equation}
where
\begin{equation}
    \Delta\mathcal{C} = \frac{D}{2} e^{-\epsilon(y_0-t_0)} \int \frac{d^4p}{(2\pi)^4}\frac{1}{E(p)}\frac{e^{-i\underline{p}.\underline{y}}e^{-ipx}e^{ip_0 t_0}}{(p_0+i\epsilon)^2-E(p)^2} \left[\frac{e^{-iE(p)(y_0-t_0)}}{p_0-i\epsilon+E(p)} -  \frac{e^{iE(p)(y_0-t_0)}}{p_0-i\epsilon-E(p)}\right] \ .
\end{equation}
Performing again the $p_0$ integral in the complex plane (close below, since $x_0\geq t_0$ by construction): 
\begin{equation}
    \int \frac{dp_0}{2\pi} \frac{e^{-ip_0(x^0-t_0)}}{[(p_0+i\epsilon)^2-E(p)^2][p_0-i\epsilon+E(p)]} =\frac{- e^{-\epsilon(x^0-t_0)}}{4 E(\underline{p})^2} \left[\frac{e^{i E(\underline{p})(x^0-t_0)}}{\epsilon}+ i\frac{e^{-i E(\underline{p})(x^0-t_0)}}{E(\underline{p})-i\epsilon}   \right]
\end{equation}
and
\begin{equation}
    \int \frac{dp_0}{2\pi} \frac{e^{-ip_0(x^0-t_0)}}{[(p_0+i\epsilon)^2-E(p)^2][p_0-i\epsilon-E(p)]} =\frac{e^{-\epsilon(x^0-t_0)}}{4 E(\underline{p})^2} \left[\frac{e^{-i E(\underline{p})(x^0-t_0)}}{\epsilon}- i\frac{e^{i E(\underline{p})(x^0-t_0)}}{E(\underline{p})+i\epsilon} .
    \right]
\end{equation}
We can then perform the angular integral and combine the results to obtain:
\begin{equation}
\begin{split}
    \Delta\mathcal{C} = -\frac{D}{8\pi^2}\frac{e^{-\epsilon(x_0+y_0-2t_0)}}{|\underline{x}-\underline{y}|} \int dp \frac{p}{E(p)^2} \left[\frac{1}{\epsilon} \cos \left( E(p)(x_0-y_0) \right) + \frac{E(p)}{(E(p)^2+\epsilon^2)} \sin \left( E(p)(x_0+y_0-2t_0) \right) \right. \\
    \left. - \frac{\epsilon}{(E(p)^2+\epsilon^2)} \cos \left( E(p)(x_0+y_0-2t_0) \right) \right] \sin\left(p |\underline{x}-\underline{y}| \right) \ .
\end{split}
\end{equation}
We can now expand in $\epsilon$ and take the limit $\epsilon\to 0$ to obtain:
\begin{equation}
    \Delta\mathcal{C} = -\frac{D}{8\pi^2} \frac{1}{|\underline{x}-\underline{y}|} \left[\left(\frac{1}{\epsilon}-(x^0+y^0-2t_0) \right)I_1 + I_3- \pi I_{\delta3}  - \pi (x^0+y^0-2t_0) I_{\delta4}\right]  \ ,
\end{equation}
where
\begin{equation}
    I_3 = \int_0^\infty dp \frac{p}{E(p)^3} \sin\left(E(p) (x^0+y^0-2t_0) \right) \sin(p|\underline{x}-\underline{y}|)
\end{equation}

\begin{equation}
    I_{\delta 3} = \int_0^\infty dp \ \delta(E(p)) \ \frac{p}{E(p)^2} \cos\left(E(p) (x^0+y^0-2t_0) \right) \sin(p|\underline{x}-\underline{y}|)= 0,
\end{equation}

\begin{equation}
    I_{\delta 4} = \int_0^\infty dp \ \delta(E(p))\frac{p}{E(p)}  \sin\left(E(p) (x^0+y^0-2t_0) \right) \sin(p|\underline{x}-\underline{y}|)=0 \ .
\end{equation}
This time around, one of the two integrals involving delta functions doesn't vanish form $m=0$, hence the Kronecker delta $\delta_{m,0}$. Note, the $I_3$ integral is essentially $I_2$ with different time components. Therefore:
\begin{equation}
    \Delta\mathcal{C} = -\frac{D}{8\pi^2} \frac{1}{|\underline{x}-\underline{y}|} \left[\left(\frac{1}{\epsilon}-(x^0+y^0-2t_0) \right)I_1 + I_3\right]  \ .
\end{equation}

\subsection*{Final result}
It is now time to add up the two contributions. As expected, the two $1/\epsilon$ divergent terms cancel, leaving a finite well-behaved 2-point function:
\begin{equation}
\begin{split}
    \mathcal{C}(x,t;t_0) &= \frac{1}{2}\mathcal{C}_\infty + \bar{\mathcal{C}} = \mathcal{C}_\infty + \Delta C \\
    & = \frac{D}{8\pi^2 } \frac{1}{|\underline{x}-\underline{y}|} \left[2(y^0-t_0) I_1 + I_2 - I_3  \right] \ .
\end{split}
\end{equation}
\subsubsection*{Massless case}
In the massless limit, the relevant integrals simplify greatly. First, consider:
\begin{equation}
\begin{split}
    \lim_{m\to 0} I_1 &= |\underline{x}-\underline{y}|\int_0^\infty dp\  \cos(p(x^0-y^0))\text{sinc}(p|\underline{x}-\underline{y}|)
    \\
    &=\frac{\pi}{4}\left[\text{sign}(x^0-y^0+|\underline{x}-\underline{y}|) - \text{sign}(x^0-y^0-|\underline{x}-\underline{y}|)\right] \\
    & = \frac{\pi}{2} \Theta(-s^2)
\end{split}
\end{equation}
where $\Theta(x)$ is the Heaviside step function and $s^2=\Delta t^2 - \Delta x^2$ is the spacetime interval. 

On the other hand:
\begin{equation}
\begin{split}
    \lim_{m\to 0} I_2 &= (x^0-y^0) |\underline{x}-\underline{y}|\int_0^\infty  dp \ \text{sinc}(p(x^0-y^0))\text{sinc}(p|\underline{x}-\underline{y}|)
    \\
    &=\frac{\pi (x^0-y^0) |\underline{x}-\underline{y}|}{|x^0-y^0+|\underline{x}-\underline{y}|| + |x^0-y^0-|\underline{x}-\underline{y}||} \\
    & = \frac{\pi}{2} |\underline{x}-\underline{y}| \Theta(s^2) + \frac{\pi}{2} (x^0-y^0) \Theta(-s^2) \ 
\end{split}
\end{equation}
and, similarly
\begin{equation}
\begin{split}
    \lim_{m\to 0} I_3 &= (x^0+y^0-2t_0) |\underline{x}-\underline{y}|\int_0^\infty dp \ \text{sinc}(p(x^0+y^0-2t_0))\text{sinc}(p|\underline{x}-\underline{y}|)
    \\
    &=\frac{\pi (x^0+y^0-2t_0) |\underline{x}-\underline{y}|}{ |x^0+y^0-2t_0+|\underline{x}-\underline{y}|| + |x^0+y^0-2t_0-|\underline{x}-\underline{y}||} \\
    & = \frac{\pi}{2} |\underline{x}-\underline{y}| \Theta(x^0+y^0-2t_0-|\underline{x}-\underline{y}|) + \frac{\pi}{2} (x^0+y^0-2t_0) \Theta(-(x^0+y^0-2t_0)+|\underline{x}-\underline{y}|) \ .
\end{split}
\end{equation}
Finally, for most applications, we will be interested in the large $y^0-t_0$ limit, we can effectively replace the latter integral by:
\begin{equation}
    \lim_{m\to 0} I_3 =\frac{\pi}{2} |\underline{x}-\underline{y}| \ .
\end{equation}

By noting that for time-like separated events $x^0+y^0-2t_0 > |\underline{x}-\underline{y}|$, we can combine all the theta-functions conditions into the following simpler result:
\begin{equation}
\begin{split}
    \mathcal{C}_0(x,y;t_0) &= \frac{1}{2}\mathcal{C}_\infty + \bar{\mathcal{C}} = \mathcal{C}_\infty + \Delta C \\
    & =  \frac{D}{16\pi } \left(\frac{(x^0+y^0-2t_0)}{|\underline{x}-\underline{y}|}-1 \right) \Theta(-s^2)\Theta(x^0+y^0-2t_0-|\underline{x}-\underline{y}|)  \ .
\end{split}
\end{equation}

\subsubsection*{Spatial gradient}
The two-point function of the spatial gradient of the field is easily computed by differentiating $\mathcal{C}$ directly, thanks to the linearity of the expectation value. By carefully handling the derivatives of the distributions, we get three terms, two of which concentrated on the lightcone (due to the derivatives of the theta function). In particular:
\begin{equation}
\begin{split}
 \mathbb{E}[\partial_i\phi(x) \partial_j\phi(y)]  =& 
      \frac{D}{16\pi} \frac{(x^0+y^0-2t_0)}{|\underline{x}-\underline{y}|}\left(\frac{\delta^i_j}{|\underline{x}-\underline{y}|^2}-\frac{3(x_i-y_i)(x_j-y_j)}{|\underline{x}-\underline{y}|^4} \right)  \Theta(-s_{xy}^2)  \\
        &+  \frac{D}{4\pi}\frac{(x^0+y^0-2t_0)}{|\underline{x}-\underline{y}|}(x_i-y_i)(x_j-y_j) \delta(-s_{xy}^2) \\
       &-\left(  \frac{D}{16\pi} \frac{(x^0+y^0-2t_0)}{|\underline{x}-\underline{y}|}- 1 \right) \left( 2\delta(-s_{xy}^2) \delta_{ij} + 4 \delta'(- s_{xy}^2) (x_i-y_i)(x_j-y_j)\right) \ ,
\end{split}
\end{equation}
where we have purposedly ignored the $\Theta$-function related to the initial conditions for simplicity, as it is irrelevant in most situations of interest -- $t_0$ is much larger than the typical scale of any Earth-based experiment. However, if we are only interested in the coincident limit (and in the sum over all directions), the solution greatly simplifies. In particular, by using the standard relation:
\begin{equation}
    \nabla^2_x\frac{1}{|\underline{x}-\underline{y}|} = -4\pi\delta^{(3)}(\underline{x}-\underline{y}) \ ,
\end{equation}
we obtain (recalling the minus sign for the spatial contractions in this signature):
\begin{equation}
\begin{split}
 \mathbb{E}[\partial_i\phi(x) \partial^i\phi(y)]  =& 
      \frac{D}{16\pi} 
      (x^0+y^0-2t_0)\left(2 \frac{\delta(-s_{xy}^2)}{|\underline{x}-\underline{y}|}-4\pi \delta^{(3)}( |\underline{x}-\underline{y}|)\Theta(-s_{xy}^2)\right)    \\
       & +\frac{D}{16\pi} \left( 6 \delta(-s_{xy}^2) + 4 |\underline{x}-\underline{y}|^2 \delta'(-s_{xy}^2) \right) \ .
\end{split}
\end{equation}
In the non-relativistic limit:
\begin{equation}
    \delta(-s^2_{xy}) \to \frac{\delta(|\underline{x}-\underline{y}|)}{|\underline{x}-\underline{y}|} \ ,
\end{equation}
and $r^2\delta'(r^2) = 0$ when integrated against any smooth test-function, whilst we always have:
\begin{equation}
    \delta^{(3)}(\underline{x}-\underline{y}) = \frac{\delta(|\underline{x}-\underline{y}|)}{4\pi|\underline{x}-\underline{y}|^2} \ .
\end{equation}
Therefore, in the non-relativistic limit we get the following covariance for the gradient of the field:
\begin{equation}
\begin{split}
 \mathbb{E}[\partial_i\phi(x) \partial^i\phi(y)]  =& 
      \frac{DT}{8\pi} 
     \frac{\delta(|\underline{x}-\underline{y}|)}{|\underline{x}-\underline{y}|^2} =   \frac{DT}{2} 
     \delta^{(3)}(\underline{x}-\underline{y})\ ,
\end{split}
\end{equation}
where we have kept only the leading term in $T\equiv-t_0$, i.e. we assume the diffusive evolution has been going on for much larger timescales than those related to any local observation.

\section{Explicit spacetime convolution for the massless field}
\label{app:spacetime_conv}
An alternative way to obtain the same result -- though less practical in the general setting -- is to perfrom the convolution:
\begin{equation}
    \mathcal{C}(x,y|t_0) = D\int_{t_0}^\infty dz^0\int d^3z\ G_R(x-z) G_R(y-z) \ ,
\end{equation}
directly by calculating the integral in the spacetime representation, where:
\begin{equation}
    G_R(x-z) = -\frac{1}{2\pi}\Theta(x^0-z^0) \delta(s_{xz}^2) = -\frac{1}{4\pi} \frac{\delta(x^0-z^0-|\underline{x}-\underline{z}|)+\delta(x^0-z^0 +|\underline{x}-\underline{z}|)}{|\underline{x}-\underline{z}|} \Theta(x^0-z^0) \ .
\end{equation}
Here we will assume $x^0 \geq y^0$. It is useful to perform the following change of variable:
\begin{equation}
    \tilde{z} = y-z \ ,
\end{equation}
trasforming the integral into:
\begin{equation}
\begin{split}
    \mathcal{C}(x,y|t_0) =\frac{D}{16\pi^2} \int^{y^0-t_0}_{-\infty}d\tilde{z}^0\int d^3\tilde{z}\ & \frac{\delta(x^0-y^0+\tilde{z}^0-|\underline{x}-\underline{y}+\underline{\tilde{z}}|)+\delta(x^0-y^0+\tilde{z}^0+|\underline{x}-\underline{y}+\underline{\tilde{z}}|)}{|\underline{x}-\underline{y}+\underline{\tilde{z}}|} \times \\
    & \qquad \times \frac{\delta(\tilde{z}^0-|\underline{\tilde{z}}|)+\delta(\tilde{z}^0+|\underline{\tilde{z}}|)}{|\underline{\tilde{z}}|} \Theta(x^0-y^0+\tilde{z}^0) \Theta(\tilde{z}^0)
\end{split}
\end{equation}

Due to the theta-function imposing positivity on $\tilde{z}^0$, the $\delta(\tilde{z}^0+|\underline{\tilde{z}}|)$ does not contribute. Integrating over $\tilde{z}^0$ we obtain:
\begin{equation}
\begin{split}
    \mathcal{C}(x,y|t_0) &= \frac{D}{16\pi^2} \int d^3\tilde{z}\  \frac{\delta(x^0-y^0+|\underline{\tilde{z}}|-|\underline{x}-\underline{y}+\underline{\tilde{z}}|)+\delta(x^0-y^0+|\underline{\tilde{z}}|+|\underline{x}-\underline{y}+\underline{\tilde{z}}|)}{|\underline{x}-\underline{y}+\underline{\tilde{z}}| |\underline{\tilde{z}}|}  \times \\
    & \hspace{200pt} \times \Theta(y^0-t_0-|\underline{\tilde{z}}|)  .
\end{split}
\end{equation}
Since $x^0 \geq y^0$, the second delta-function is irrelevant. The theta-function imposes the restriction on $|\underline{\tilde{z}}|$ due to the finite evolution in time. It is now convenient to perform the spatial integral in spherical polars $(\tilde{z},\theta,\phi)$, where the role of the unit $\hat{k}$ vector with respect to which the angles are defined is played by $(\underline{x}-\underline{y})/|\underline{x}-\underline{y}|$. Then, using:
\begin{equation}
    |\underline{x}-\underline{y}+\underline{\tilde{z}}| = \sqrt{|\underline{x}-\underline{y}|^2+\tilde{z}^2 + 2|\underline{x}-\underline{y}|\tilde{z} \cos\theta} \ ,
\end{equation}
meaning that the delta-function condition is satisfied for $\theta_*$ s.t.:
\begin{equation}
    \cos \theta_* = \frac{(x^0-y^0)^2-|\underline{x}-\underline{y}|^2}{2 \tilde{z}|\underline{x}-\underline{y}|} + \frac{x^0-y^0}{|\underline{x}-\underline{y}|}
\end{equation}

This clearly implies that $x$ and $y$ must be spacelike separated spacetime events -- if timelike the RHS is larger than 1 (recall that $x^0-y^0 \geq 0$). In terms of $\cos \theta$, the delta-function can be expressed as:
\begin{equation}
\begin{split}
    \delta(x^0-y^0+|\underline{\tilde{z}}|-|\underline{x}-\underline{y}+\underline{\tilde{z}}|) &=  \frac{\sqrt{|\underline{x}-\underline{y}|^2+\tilde{z}^2 + 2|\underline{x}-\underline{y}|\tilde{z} \cos\theta_*}}{\tilde{z}|\underline{x}-\underline{y}|}\delta(\cos\theta-\cos\theta_*) \times \\
    &\hspace{100pt} \times \Theta(-(x^0-y^0-|\underline{x}-\underline{y}|)) \ .
\end{split}
\end{equation}
Now, in spherical polars the integral becomes simply (ignoring the theta-function for brevity):
\begin{equation}
\begin{split}
    \mathcal{C}(x,y|t_0) &=\frac{D}{16\pi^2|\underline{x}-\underline{y}|} \int_0^{2\pi}d\phi \int_{-1}^{1} d(\cos\theta)\int_0^{y^0-t_0} d\tilde{z}  \frac{\sqrt{|\underline{x}-\underline{y}|^2+\tilde{z}^2 + 2|\underline{x}-\underline{y}|\tilde{z} \cos\theta_*}}{\sqrt{|\underline{x}-\underline{y}|^2+\tilde{z}^2 + 2|\underline{x}-\underline{y}|\tilde{z} \cos\theta}} \times \\
    &\hspace{200pt} \times\delta(\cos\theta-\cos\theta_*) \ .
\end{split}
\end{equation}
Meaning that the final result is:
\begin{equation}
    \mathcal{C}(x,y|t_0) =\frac{D}{16\pi } \left( \frac{y^0+x^0-2t_0}{|\underline{x}-\underline{y}|} -1\right)\Theta(-s_{xy}^2) \Theta(x^0+y^0-2t_0- |\underline{x}-\underline{y}|)\ ,
\end{equation}
which agrees with the result found from the momentum representation.

\section{Large forces on test masses in a stochastic Yukawa theory} \label{app:forces}
In this appendix we discuss how fluctuations of the scalar field affects the motion a test particles interacting with it as if it were a potential -- we are treating here the stochastic scalar as a toy model. 
Whether the quantum fluctuations of a quantum field induce stochastic motion of a particle (sourcing and responding to the field) has been addressed in multiple studies~\cite{gour1999will,yu2004vacuum,johnson2002stochastic,parikh2020the-noise}. The result is highly sensitive on whether the field is in the vacuum or thermal state, and on the localisation of the particle interacting with the field. However, generally the effect is not large enough to be measurable when the field is in the vacuum state and becomes important only at high temperatures~\cite{johnson2002stochastic,parikh2020the-noise,carney2024response}. As we see now, the converse is true for the classical stochastic fluctuations.

For a back-of-the-envelope estimation of the forces acting solid extended objects, let's treat the test mass as a classical constant density $\rho$ sphere of diameter $R$. In the case of quantum particles, we will consider them localised within their Compton wavelength, i.e. $R=\lambda_c/2$. Then, in the Yukawa model, the particle responds to spatial gradients of the field. In particular, ignoring backreaction effects (i.e. radiation-reaction forces), the force on the test particle is simply given by:
\begin{equation}
    F_i = -\rho \int d^3x  \ \Theta \left(R-r \right) \partial_i\phi\ ,
\end{equation}
where $r$ is the radial coordinate of the 3D cartesian system with origin at the center of the spherical mass. On average, $\phi$ vanishes, and so does $\partial_i \phi$, meaning that the force on the particle induced by the stochastic fluctuations is zero on expectation in any specific direction. However, the norm of the vector itself has non-zero expectation. Indeed:
\begin{equation}
    \mathbb{E}[F^2] = \mathbb{E}[F_iF^i] = \rho^2 \int d^3x \int d^3 y \ \Theta \left(R-r_x \right) \Theta \left(R-r_y \right) \mathbb{E}[\partial_i\phi \partial_j\phi] 
\end{equation}
Extracting the two-point function of the spatial gradient of the field is straightforward -- it suffices to take the spatial gradient of $\mathcal{C}$ by linearity of the expectation value. The complete expression is cumbersome -- see Appendix~\ref{app:four_int}. For simplicity, we assume the timescale of the experiment is much shorter than the total time of the evolution of the system $T$, meaning $x^0+y^0-2t_0 \approx 2T$. In the non-relativistic limit, the expression simplifies to:
\begin{equation}
    \mathbb{E}[\partial_i\phi(x) \partial^i\phi(y)]  = 
      \frac{DT}{2} 
     \delta^{(3)}(\underline{x}-\underline{y}) .
\label{eq:twopoint}\end{equation}
Again, this maps nicely to the equilibrium 2-point function, Equation \eqref{eq:deldel_th}.  Then:
\begin{equation}
    \mathbb{E}[F^2] = c^5D T \rho^2 V = \frac{DTM^2c^5}{V}\ ,
\end{equation}
where $M$ is the mass of the particle, $V$ its volume and we have re-introduced factors of $c$ (and $G$, which does not appear as it is already implicitly included in the dimensionless diffusion coefficient). As measured by a device which coarse-grains the observation on a time-scale $t_c$, the centre of mass of the particle evolves as:
\begin{equation}
    M \ddot{z}^i = F^i + f^i(t) \ ,
\end{equation}
where $\underline{F}$ is the sum of any external force acting on the particle, whilst $f^i(t)$ is a stochastic force obeying:
\begin{equation}
    \mathbb{E}[f^i(t)] = 0 \ , \qquad \mathbb{E}[f^i (t)f^j(t')] = \frac{c^5 M^2DT}{4\pi R^3} t_c \ \delta^{ij} \delta(t-t') \ ,
\end{equation}
where we have defined $T=0$ to be $t_0$, the initial time of diffusion of the scalar field, and have weighted the $\delta$-function by $t_c$, the time scale associated with the spatial averaging.

First, let's consider the potential effect this would have had on one of the  $M\approx1 \ \mathrm{Kg}$ free-falling test masses in the LISA Pathfinder Technololy Package, the ESA technology demonstration mission for the future gravitational wave detector LISA. The test masses used were objects of radius $R\approx 5 \times 10^{-2} \ \mathrm{m}$~\cite{armano2019lisa}. LISA Pathfinder results are quoted with respect to the variance in the relative acceleration spectral density of the masses -- where the maximum frequency to which the experiment is sensitive is $\omega_* = 1/t_c \approx 1 \ \mathrm{Hz}$. Being the effective force $\delta$-correlated in time and space, the two masses are acted upon by uncorrelated random forces -- and the spectral density of the acceleration variance of the test mass has flat frequency profile of size. Then:
\begin{equation}
    \sigma_{aa}^2(\omega) = D \times 10^{62}\ \mathrm{m}^2 \mathrm{s}^{-4}\mathrm{Hz} \ \lessapprox  10^{-30} \ \mathrm{m}^2 \mathrm{s}^{-4}\mathrm{Hz} \ .
\end{equation}
Under this model, the accumulation over cosmological times of scalar stochastic wave would provide a formidable bound on the diffusion coefficient of the CQ theory:
 \begin{equation}
     D \lessapprox 10 ^{-92} \ .
 \end{equation}
Coherence experiments on spatial superpositions currently lower bound the dimensionless diffusion coefficient by $D\gtrapprox 10^{-71}$~\cite{grudka2024renormalisation} by the decoherence-diffusion trade-off, meaning that the model would be ruled out by experiments. However, as mentioned before, these back-of-the-envelope estimation of the induced force by the stochastic fluctuations is not to be taken at face value as a good proxy for the predictions on tabletop experiments -- both because scalar waves are not necessarily expected in CQ gravity, and because the linearised model cannot be trusted at short scales, where its irregular solution can in principle be cured by the self-interaction of the gravitational field. Yet, it is a strong indication that, unless renormalisation greatly reduces the size of the fluctuations in the UV by several orders of magnitude, CQ gravity can be experimentally tested with current technology.

The force induced by the fluctuations can be very large indeed. Another way to see why CQ gravity is in danger of being falsified unless the stochastic fluctuations are greatly reduced in the UV, is to consider their effects on subatomic particles. For example, consider the force on an electron-sized particle:
\begin{equation}
    \bar{F} \approx \sqrt{D} \times10^{19} \ \text{N} \ ,
\end{equation}
which can be used set an order of magnitude bound on $D$. A good benchmark is the stability of Rydberg atoms, atoms in which the outermost electron is pushed to very excited radial states (i.e. the principal quantum number of the radial wavefunction $n$ is very large, up to $n\approx700$). In these case, the force that keeps the electron bound is of the order:
\begin{equation}
    F_R \approx 10^{-19} \ \text{N} \ ,
\end{equation}
meaning that, for these atoms to exist, the diffusion coefficient needs to be extremely small, i.e. $D \lessapprox 10^{-76}$ -- still violating the decoherence-diffusion trade-off. If the diffusion coefficient were larger than this, the nucleus and the elctrons would each feel a force in an uncorrelated direction of the typical size of the force that keeps the electron bound, meaning that the latter would be stripped out of its orbit immediately.

Of course, it is important to caveat this discussion once more. The KG stochastic field is in no way a good model of CQ gravity. As soon as these fluctuations become larger than the background gravitational field, the linear approximation fails and the model becomes unpredictive. Hence, we cannot use it to reliably place any bounds of the sort: we anticipate that this unbounded diffusion will be cured once the non-linear character of the gravitational field is considered -- much like non-linearities in the stochastic wave equation allow for the existence of regular solutions~\cite{conus2008the-non-linear}. However, this back of the envelope calculation is still useful to showcase how CQ gravity can produce strong experimental predictions, opening the way to novel methods to test the quantum nature of gravity.

\bibliography{comprehensive}

\end{document}